\documentclass[aps,twocolumn,prd,superscriptaddress,showpacs,preprintnumbers,
amsmath,amssymb,nofootinbib] 
{revtex4-1}

\usepackage{color}
\usepackage{graphicx}
\usepackage{amsmath, amssymb}
\usepackage{amsfonts}
\usepackage{dcolumn}
\usepackage[caption=false]{subfig}
\usepackage{hyperref}

\usepackage[english]{babel}
\usepackage{comment}
\usepackage{xspace} 
\usepackage{mathtools} 

\newcommand{\tr}{\ensuremath{\operatorname{tr}}}
\newcommand{\Omegaqq}{\ensuremath{\Omega_{\bar{q}q}}}
\newcommand{\vev}[1]{\ensuremath{\left\langle #1 \right\rangle}}

\def\eqref#1{(\ref{#1})}
\def\fig#1{Fig.~\ref{#1}}

\newcommand{\Phibar}{\ensuremath{\bar{\Phi}}}
\newcommand{\LPQM}{\ensuremath{\mathcal{L}_{\textrm{PQM}}}\xspace}

\def\Dr{{D\!\llap{/}}\,}

\def\0#1#2{\frac{#1}{#2}}

\def\qbar{\bar{q}}

\newcommand*{\defeq}{\vcentcolon=}

\graphicspath{{../figures/}{./}{./figures/}}

\begin{document}

\title{Pion and $\eta$-meson mass splitting at the two-flavour chiral crossover}

\author{Markus Heller}
\affiliation{
Dip. di Fisica "Galileo Galilei",
Via Marzolo 8,
I-35131 Padova, Italy
}
\affiliation{Institut f\"ur Theoretische Physik,
Universit\"at Heidelberg, 
Philosophenweg 16, D-69120 Heidelberg, Germany}

\author{Mario Mitter}
\affiliation{Institut f\"ur Theoretische Physik, Universit\"at Heidelberg, 
Philosophenweg 16, D-69120 Heidelberg, Germany}


\begin{abstract}
  \noindent We study the splitting in the screening mass of pions and the $\eta$-meson across the chiral crossover.
	    This splitting is determined by the 't~Hooft determinant. We use results for the renormalisation group scale 
	    dependence of the 't~Hooft determinant obtained within the functional renomalisation group in quenched QCD with
	    two flavours. The scale dependence of the 't~Hooft determinant is mapped to its temperature dependence with the help
	    of a Polyakov-quark-meson model. As a result we obtain the temperature dependence of the splitting in the 
	    screening mass of pions and the $\eta$-meson.
\end{abstract}

\maketitle

\section{Introduction}\label{sec:intro}

The axial $U(1)_A$-symmetry of Quantum Chromodynamics is broken by a quantum
anomaly. As a consequence the pseudoscalar singlet meson does not appear as a massless mode 
in the spectrum of QCD in the phase of spontaneously broken 
chiral symmetry \cite{Witten:1979vv,Veneziano:1979ec}. This phenomenon 
can also be understood in terms of the 't~Hooft determinant \cite{'tHooft:1976up,'tHooft:1976fv}.
It is a $U(1)_A$-symmetry breaking $2N_f$-quark interaction that gets contributions from
fluctuations in the topological charge. 
Recently, non-perturbative results for the four-quark interaction channels, 
and in particular for the 't~Hooft determinant, have become available from investigations of quenched
continuum QCD with two flavours \cite{Mitter:2014wpa}, see also \cite{Pawlowski:1996ch}.

At temperatures of roughly $150-160$ MeV, QCD with $2+1$ flavours experiences a rapid crossover to a phase that approximately respects
chiral symmetry and breaks center symmetry \cite{Aoki:2006we,Aoki:2009sc,Bhattacharya:2014ara}. Although the situation is less clear at finite chemical potential,
qualitative changes are expected also at large densities, see e.g. 
\cite{Buballa:2003qv,Schaefer:2006sr,Alford:2007xm,Fukushima:2010bq,Fukushima:2013rx,Pawlowski:2014aha} for reviews. 
At temperatures far above the crossover and also at large densities, 
the effects of the axial anomaly are expected to vanish, since topological charge fluctuations become suppressed 
\cite{Gross:1980br,Schafer:1996wv}. Furthermore, it has been argued that the splitting between the mass of the pseudoscalar singlet meson and 
the pion could become small immediately above the chiral crossover \cite{Kapusta:1995ww}. Experimental signs for
such an effective restoration of the $U(1)_A$-symmetry have been found by \cite{Csorgo:2009pa,Vertesi:2009wf},
who report a drop in the in-medium mass of the pseudoscalar singlet meson of at least $200$ MeV. 

A restoration of the $U(1)_A$-symmetry would have a qualitative impact on the nature of the phase transition
in the two-flavour chiral limit. In the presence of the axial anomaly in terms of the 't~Hooft determinant this transition
is expected to be of second order in the $O(4)$ universality class \cite{Pisarski:1983ms}. If, however, the 
$U(1)_A$-symmetry were restored at the chiral transition, a first order transition or a second order transition
in the universality class of $U(2)_L\times U(2)_R/U(2)_A$ 
could take place \cite{Butti:2003nu,Pelissetto:2013hqa,Meggiolaro:2013swa,Grahl:2013pba,Grahl:2014fna}.

The possibility of an effective restoration of the $U(1)_A$-symmetry has been addressed in several
lattice QCD simulations. A degeneracy in the correlators of pion and pseudoscalar singlet meson has been observed 
in the chirally symmetric phase in a two-flavour simulation with overlap and domain wall fermions
 \cite{Cossu:2013uua,Cossu:2015lnb}. 
On the other hand, \cite{Bazavov:2012qja,Bhattacharya:2014ara} find effective restoration
only at larger temperatures of $196$ MeV with $2+1$ flavours of domain wall fermions
and \cite{Sharma:2013nva,Dick:2015twa}, using highly improved staggered fermions,
do not see it even at $1.5$ times the crossover temperature.
The phenomenological implications of the axial anomaly and different scenarios for its fate at the chiral crossover have been 
investigated with a Dyson-Schwinger approach using models for the quark-gluon interaction \cite{vonSmekal:1997dq,Alkofer:2008et,Benic:2011fv,Benic:2014mha}, 
in the Nambu--Jona-Lasinio (NJL) model \cite{Chen:2009gv,Zhang:2011xi,Powell:2011ig,Bratovic:2012qs} 
and with quark-meson \cite{Schaefer:2008hk,Mitter:2013fxa,Herbst:2013ufa} as well as linear sigma models \cite{Fejos:2015xca,Eser:2015pka}.

In this work we use results for the energy-momentum scale dependence of the 't~Hooft determinant
\cite{Mitter:2014wpa}, where the functional renormalisation group (RG) in Wetterich's formulation \cite{Wetterich:1992yh}
has been used to calculate the effective action of quenched QCD with two flavours. From this functional renormalisation group, 
a coupled set of equations for the $1$PI correlation functions, similar to the Dyson-Schwinger equations, can be derived, see e.g. 
\cite{Berges:2000ew,Roberts:2000aa,Alkofer:2000wg,Pawlowski:2005xe,Fischer:2006ub,Gies:2006wv,Binosi:2009qm,Braun:2011pp} for reviews.
Additionally, we use the Polyakov-quark-meson (PQM) model \cite{Roessner:2006xn,Schaefer:2007pw,Herbst:2010rf} with two flavours as a qualitative
description of the chiral crossover. With the help of this model we derive a mapping of the renormalisation group scale dependence of the 't~Hooft determinant
to its temperature dependence for the investigation of the mass splitting between $\eta$-meson and pion at the chiral crossover.

This paper is organised as follows: In Section \ref{sec:PQM2p1} we briefly discuss the PQM
model with two quark flavours and its behaviour at the chiral crossover. Section \ref{sec:thooftdet} discusses
the calculation of the 't~Hooft determinant \cite{Mitter:2014wpa} and the derivations of its temperature dependence. 
In Section \ref{sec:results} we discuss our main result, the mass splitting
between $\eta$-meson and pion at the chiral crossover and we summarise and conclude in \ref{sec:conclusion}.

\section{Polyakov-quark-meson model}\label{sec:PQM2p1}

\subsection{Lagrangian}

The Euclidean Lagrangian of the 2-flavour Polyakov-quark-meson
model with axial symmetry breaking term has the form \cite{Schaefer:2007pw}
\begin{align}
		\LPQM =& \qbar \left[ \Dr + \frac{h_{\pi}}{2} \left( \sigma + i\gamma^5\vec{\tau}\vec{\pi} \right) + \frac{h_{\eta}}{2} \left( \vec{\tau}\vec{a} + i\gamma^5\eta \right) \right]q \nonumber\\
			&+ \tr\left(\partial_{\mu}\Sigma\partial^{\mu}\Sigma^{\dagger}\right) + U(\rho, \xi) + \mathcal{U}(\Phi, \Phibar) \;,	\label{eq:LPQM}
\end{align}
with the meson field $2\Sigma = (\sigma + i\eta) + (\vec{a} + i\vec{\pi})\vec{\tau}$ and Pauli matrices $\vec\tau$. The mesonic
potential
\begin{align}
	U(\rho, \xi) = m_{\rho}^2\, \rho + m_{\xi}^2\, \xi + g\, \rho^2 - c\, \sigma \;,
	\label{eq:mesonPot}
\end{align}
is a function of the chirally symmetric operators $\rho= \tr\left( \Sigma \Sigma^{\dagger} \right)$
and the Kobayashi-Maskawa-'t~Hooft determinant $\xi = \det\Sigma + \det\Sigma^{\dagger}$ \cite{Kobayashi:1970ji,'tHooft:1976up,'tHooft:1976fv}. 
The latter breaks the $U(1)_A$-symmetry and leads to a mass splitting between the ($\sigma$-$\vec\pi$)- and ($\eta$-$\vec a$)-mesons.
Additionally, an explicit symmetry breaking term $c$ appears in this potentials which mimics
a non-vanishing current quark mass.

In the PQM-model the covariant derivative $\Dr = \gamma^{\mu}(\partial_{\mu} - iA_0\gamma^0)$ depends only on the non-fluctuating background field $A_0$. 
Therefore also the Polyakov loop, given as the thermal expectation value of the path-ordered and colour-traced Wilson loop,
\begin{align}
	\Phi = \frac{1}{N_c}\left\langle\tr_c\mathcal{P}\exp\left[{-i \int_{0}^{\beta}\!\textrm{d}\tau A_0(\vec{x}, \tau) }\right]\right\rangle_{\beta=1/T}  \;,
	\label{eq:loopVar}
\end{align}
depends only on this constant background.
To provide the gluonic background we use a polynomial ansatz for the effective Polyakov loop potential $\mathcal{U}(\Phi,\Phibar) $ \cite{Roessner:2006xn}
with the same parameters as \cite{Herbst:2010rf}.
Since we consider only the case of vanishing chemical potential, we have $\Phibar = \Phi$, and the potential $\mathcal{U}(\Phi,\Phibar) $ is therefore
a function of $\Phi$ alone.

\subsection{Effective potential}

\begin{figure}[t]
	    \centering
	    \includegraphics[width=1\linewidth]{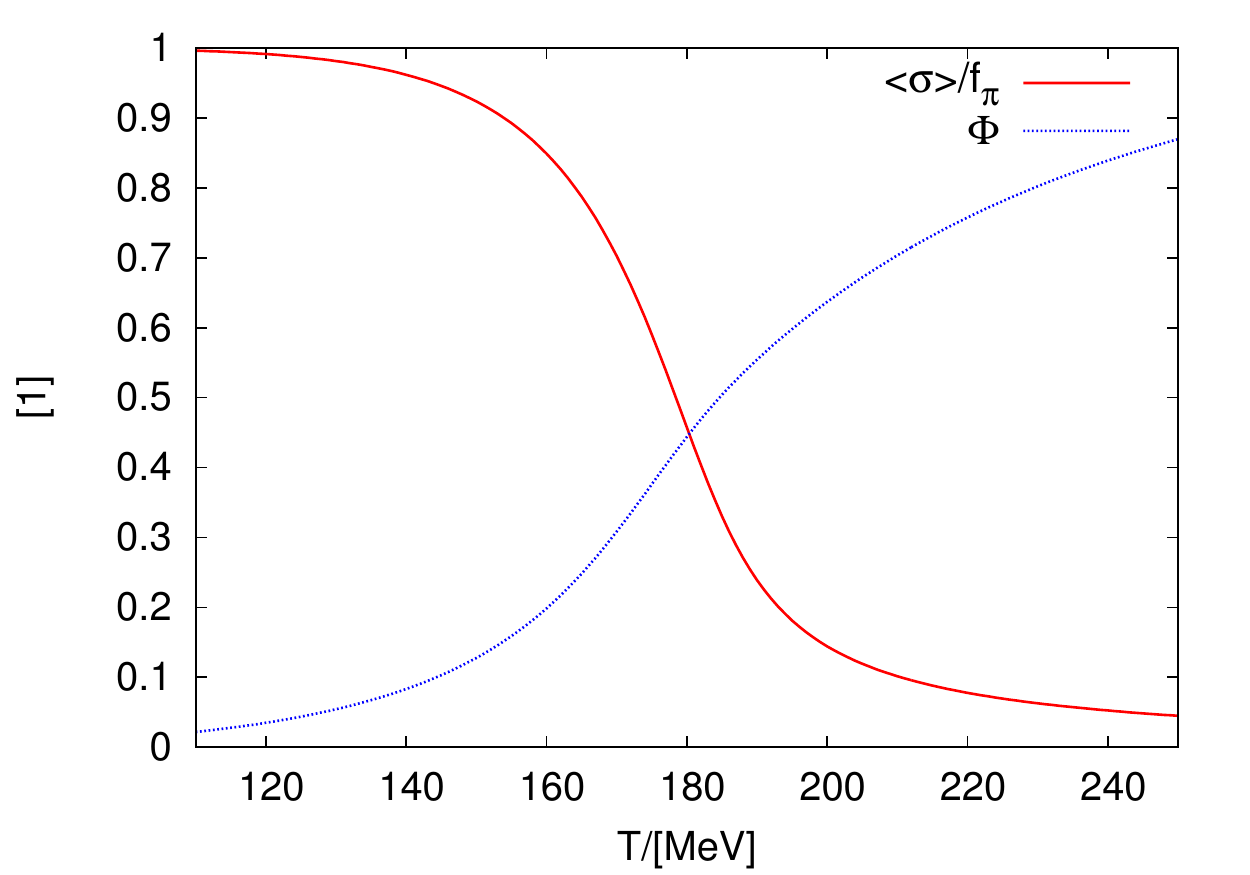}
	    \caption{Order parameters for the chiral, $\langle\sigma\rangle/f_\pi$, and deconfinement transition, $\Phi$, as functions of the temperature $T$.} 
	    \label{fig:orderparams}      
\end{figure}

\begin{figure*}
	\centering
	\subfloat[Comparison of the strength of the pion and $\eta$-meson channel in the four-quark interaction.]{\includegraphics[width=0.5\linewidth]{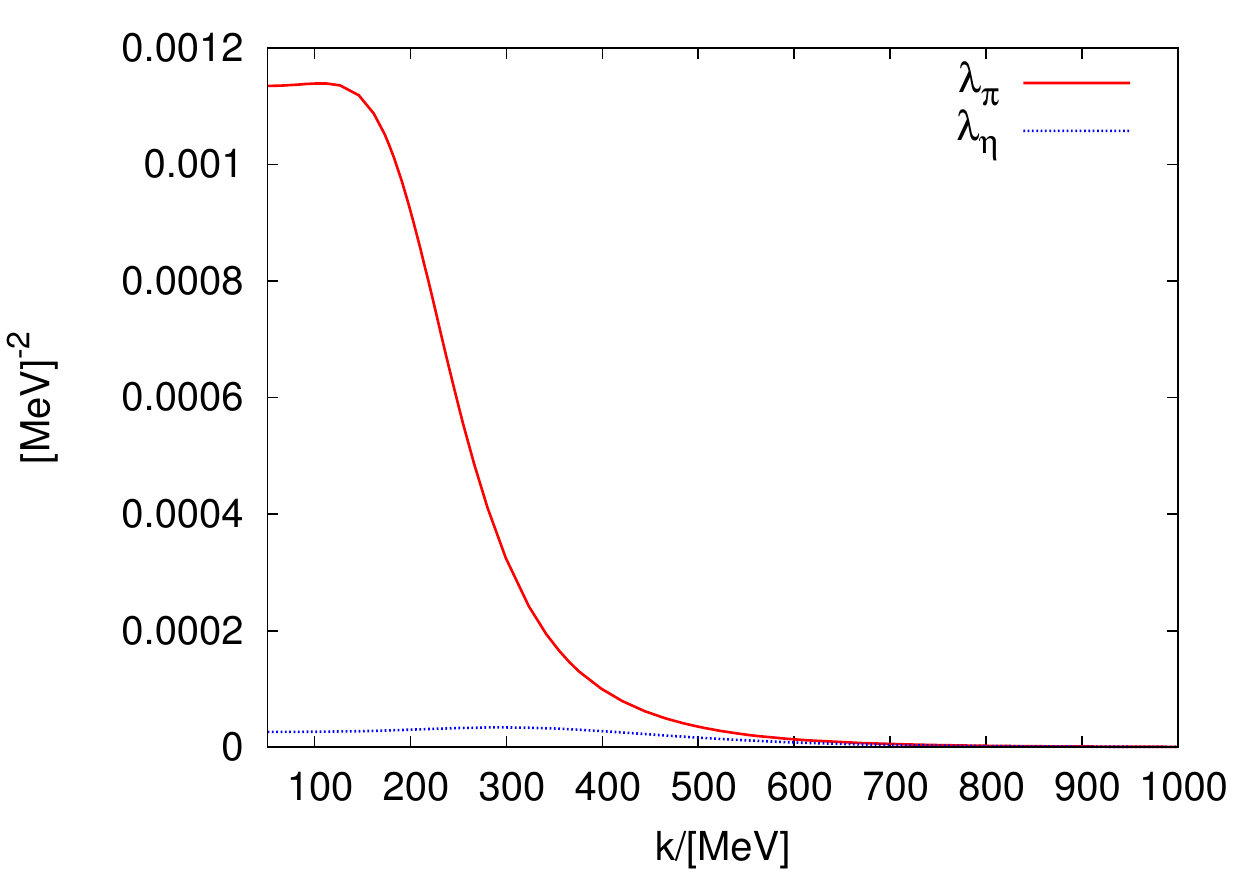}}
	\subfloat[Comparison of the symmetric coupling  $2\lambda_{(S-P)_+}=\lambda_\pi+\lambda_\eta$ with the 't~Hooft determinant coupling $2\Delta=\lambda_\pi-\lambda_\eta$.]{\includegraphics[width=0.5\linewidth]{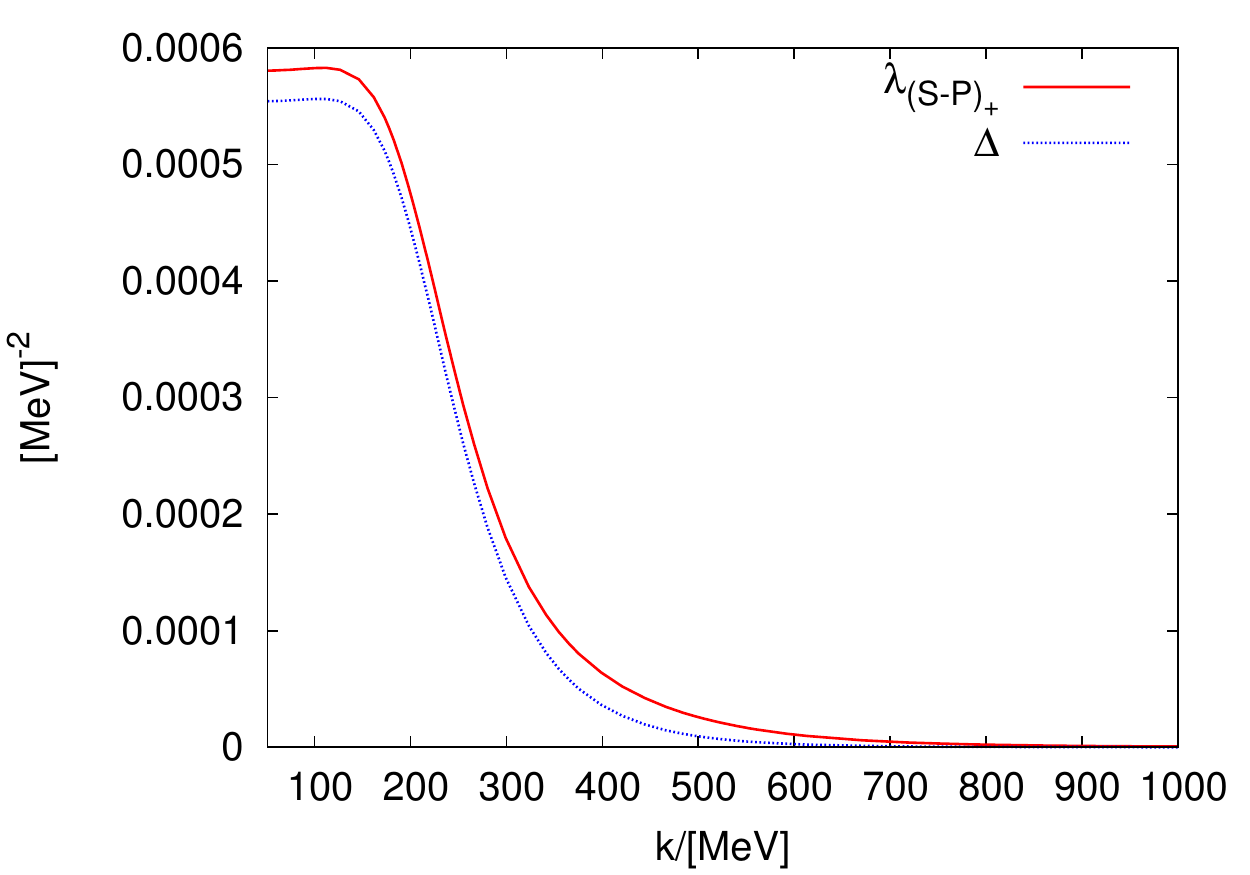}}
	\caption{Four-quark interactions as function of renormalisation group scale $k$ \cite{Mitter:2014wpa}.}
	\label{fig:fourfermi}
\end{figure*}

We calculate the effective potential $\Omega_{MF}$ at finite temperature in the extended mean-field approximation \cite{Skokov:2010sf} which ignores all mesonic fluctuations. 
The remaining path integral is Gaussian, leading to the determinant of the fermionic kinetic operator. It depends on the Yukawa interactions which we 
assume to be degenerate $h\equiv h_{\pi} = h_{\eta}$. Using $\Phi = \Phibar$ at $\mu = 0$ the result is
\begin{align}
	\Omega_{MF} &= \Omegaqq + U(\rho, \xi) + \mathcal{U}(\Phi)\; , \label{eq:Omega}
\end{align}
where
\begin{align}
		\Omegaqq &= -12\int^\Lambda \!\frac{\textrm{d}^3\!p}{(2\pi)^3} E_q - 8T\int \!\frac{\textrm{d}^3\!p}{(2\pi)^3} \\
		&\log\left[ 1 + 3\Phi e^{-\beta E_q} + 3 \Phi e^{-2\beta E_q} + e^{-3\beta E_q}  \right] \nonumber\; .
\end{align}
Here, $E_q^2 = \vec{p}^2 + \frac{h^2}{2}\rho$ is the quark energy. The potential is regularised
with a momentum cutoff $\Lambda$ and then renormalised such that the correct value for the pion mass and pion decay constant
$\vev{\sigma}=f_{\pi}$ are reproduced at $T=0$. The temperature dependence of the 't Hooft determinant in this setting is 
trivial since the quark loop, being a function of $\rho$ only, does not induce any anomalous contributions. 
Thus, the $\eta$-mass would show only a trivial temperature dependence which is proportional to the change in the pion mass
in this effective description.

\subsection{Crossover}

Once the model parameters have been fixed to yield physical values for the observables, the Polyakov-quark-meson model 
gives a qualitative description for the QCD crossover at finite temperature. 
To demonstrate this, we show the order parameters in \fig{fig:orderparams}. The normalised chiral condensate 
shows a rapid change to a very small value around temperatures of $170$ to $180$ MeV, which is the usual
value found in these type of models for the temperature of the chiral crossover.
The Polyakov loop, $\Phi$, increases at the same temperatures, indicating
a crossover to a deconfined phase of broken center symmetry.
When comparing to results from lattice QCD \cite{Aoki:2009sc,Bhattacharya:2014ara}, the crossover temperature in this model
is too large by about $20$ MeV. 
To some extent this can be remedied by improving the parameterisation of the Polyakov loop
Potential and adding a strange quark \cite{Haas:2013qwp,Herbst:2013ufa}, but it is still an indication for the qualitative
nature of the description provided by the PQM-model. 
To improve its applicability, this model can be embedded in a full QCD calculation via the dynamical hadronisation technique 
\cite{Gies:2001nw,Pawlowski:2005xe,Floerchinger:2009uf}. Such an approach has been used e.g. quantitatively in 
quenched QCD in \cite{Mitter:2014wpa} and qualitatively also for unquenched QCD in \cite{Braun:2014ata}.

\section{'t Hooft determinant}\label{sec:thooftdet}

We use results for the 't Hooft determinant from a calculation within quenched QCD with two quark flavours \cite{Mitter:2014wpa}.
This approach uses only the strong coupling strength and bare quark mass as input at 
perturbative momentum scales. From this input, the effective action $\Gamma[\phi]$ 
is calculated in a vertex expansion by solving the Wetterich equation \cite{Wetterich:1992yh}
\begin{eqnarray}\label{WetterichEq}
 \partial_k\Gamma_k[\phi] = \frac{1}{2}\text{Tr}\left[\frac{1}{\Gamma_k^{(2)}+R_k}\,\partial_kR_k\right]\ .
\end{eqnarray}
By the integration of infinitesimal momentum shells, the resulting trajectory for the effective average
action, $\Gamma_k[\phi]$, connects the renomalised perturbative 
action $S[\phi]=\lim_{k\rightarrow\Lambda}\Gamma_k[\phi]$, defined in terms of strong coupling strength and 
bare quark mass at some large momentum $\Lambda$, with the full quantum effective action 
$\Gamma[\phi]=\lim_{k\rightarrow0}\Gamma_k[\phi]$. The momentum-shell integration is controlled by the 
regulator function $R_k$ which acts as a momentum dependent mass-term.

As the momentum shell integration approaches non-perturbative momenta below $\mathcal{O}$($1$) GeV, four-quark
interactions are created via two-gluon exchange diagrams, e.g. \cite{Braun:2011pp}. If these four-quark interactions
are approximated by a momentum-independent coupling constant, chiral symmetry breaking is signalled by 
a singularity in this coupling. This divergence is as consequence of the emergence of the pion
pole in the spontaneously broken phase. To avoid this divergence, one has to include momentum dependencies
in the four-quark interaction. One possibility to do so with manageable effort is to use the dynamical hadronisation technique
\cite{Gies:2001nw,Pawlowski:2005xe,Floerchinger:2009uf}, see also \cite{Mitter:2014wpa,Braun:2014ata,Rennecke:2015eba} for recent applications. 
In each momentum shell integration step, this technique rewrites the four-quark interactions in terms of meson exchange. 
This naturally includes the correct momentum dependence close to the pion pole, thus avoiding any singularities.

\subsection{Renormalisation group scale dependence of the 't Hooft determinant}\label{sec:thooftdet_k}

To be able to identify the resonant four-quark interaction channel a fierz-complete basis with ten
basis elements has been used in \cite{Mitter:2014wpa}. The momentum dependence of the resonant
pion channel has been taken into account via the dynamical hadronisation technique.
The choice of only hadronising this single channel uses the knowledge that the $U(1)_A$-anomaly
will break the symmetry between pions and $\eta$-meson. 
A more thorough investigation of the implications of this choice and the $U(1))_A$-anomaly
will be presented elsewhere.

Here we are especially interested in the
four-quark channel corresponding to the exchange of the $\eta$-meson (and also $\vec{a}$-meson) and its difference to
the pion-channel, which is directly proportional to the 't~Hooft determinant in the case of two quark flavours.
The corresponding results for these two channels are shown in Fig.~\ref{fig:fourfermi}. 
These results have been obtained in the truncation of \cite{Mitter:2014wpa} and have been rescaled such that a 
unit residue at the pion pole is guaranteed.
Additionally the wave function renormalisations of pions and 
$\eta$-meson have been assumed to be degenerate. We clearly see that the four-quark interactions reach a considerable strength only
at non-perturbative momenta. In this regime, the 't~Hooft determinant is almost as strong as the symmetric four-quark channel. 
As a consequence, the pion channel becomes dominant, wherase the $\eta$-meson is comparably heavy.
Although hardly visible in Fig.~\ref{fig:fourfermi}, the $\eta$-pion splitting is already present at scales above the chiral symmetry
breaking scale. One possible source for this splitting is the previously mentioned asymmetric choice of 
only hadronising the pion-$\sigma$-meson channel.

\subsection{Temperature dependence of the 't Hooft determinant}\label{sec:thooftdet_T}

As discussed already, the quark-meson model can be derived from QCD within the functional renormalisation group approach via the 
dynamical hadronisation technique.
Within this approach, mesons are introduced as auxiliary fields. The momentum dependence of the dynamically 
created four-quark interactions is then rewritten in terms of the exchange of meson $\phi$, i.e.
\begin{align}
 \Gamma_{(\bar q q)^2}^{(4)}\longrightarrow \Gamma_{(\bar q q)\phi}^{(3)}(\Gamma_{\phi\phi}^{(2)})^{-1}\Gamma_{(\bar q q)\phi}^{(3)}\ .
\end{align}
The couplings of the four-quark channels corresponding to pion and $\eta$-meson exchange, $\lambda_\pi$ and $\lambda_{\eta}$, 
are therefore related to the corresponding Yukawa interactions and meson masses via the relations
\begin{align}
	\lambda_{\pi} = \frac{h_{\pi}^2}{2m_{\pi}^2}  \qquad
	\lambda_{\eta} = \frac{h_{\eta}^2}{2m_{\eta}^2} \;,
	\label{eq:lambda}
\end{align}
which hold at each renormalisation group scale $k$.

For calculating the temperature dependence of the $\eta$-meson mass we use the temperature independent approximation 
$h \equiv h_{\pi} = h_{\eta}$. For given $h$, the temperature dependence of the meson masses, $m_\phi(T)$, is then directly
proportional to temperature dependence of the corresponding four-quark interaction strength, $\lambda_\phi(T)$.
Consequently, we are left with calculating the temperature dependence of $\lambda_\eta(T)$. To do so, we assume that
the temperature dependence of a given coupling strength can be mapped onto its renormalisation group scale dependence,
since both quantities are energy-momentum scales. A similar approximation has been used successfully for obtaining the 
strong running coupling as a function of temperature and magnetic field in \cite{Mueller:2015fka}.
To calculate the precise relation between the temperature and RG-scale dependence we use the temperature dependence
of the pion mass, $m_{\pi,\text{PQM}}(T)$, as obtained within the PQM model. The pion mass gives us the corresponding four-quark interaction 
strength $\lambda_{\pi,\text{PQM}}(T)$ via (\ref{eq:lambda}) with $h_\pi=h$.
Finally, we use the corresponding coupling
$\lambda_{\pi,\text{QCD}}(k)$, obtained in the approach presented in \cite{Mitter:2014wpa}, to relate the RG-scale $k$
with the temperature $T$ by demanding $\lambda_{\pi,\text{PQM}}(T)\equiv \lambda_{\pi,\text{QCD}}(k)$. This condition
maps each temperature $T$ to a corresponding RG scale $k(T)$ and we obtain finally
\begin{align}
	m_{\eta}^2(T) = \frac{h^2}{2}\left( \frac{h^2}{2m_{\pi,\text{PQM}}^2(T)} - 2\Delta_{\text{QCD}}(k(T)) \right)^{-1}\; ,
	\label{eq:meta}
\end{align}
from the RG-running of the 't~Hooft determinant
\begin{align}
	\Delta_{\text{QCD}}(k) \defeq \frac{\left( \lambda_{\pi,\text{QCD}}(k) - \lambda_{\eta,\text{QCD}}(k) \right)}{2}\; .
	\label{eq:delta}
\end{align}

\section{$\eta$- and $\pi$-meson mass splitting at chiral crossover}\label{sec:results}

\begin{figure}[t!]
	    \centering
	    \includegraphics[width=1\linewidth]{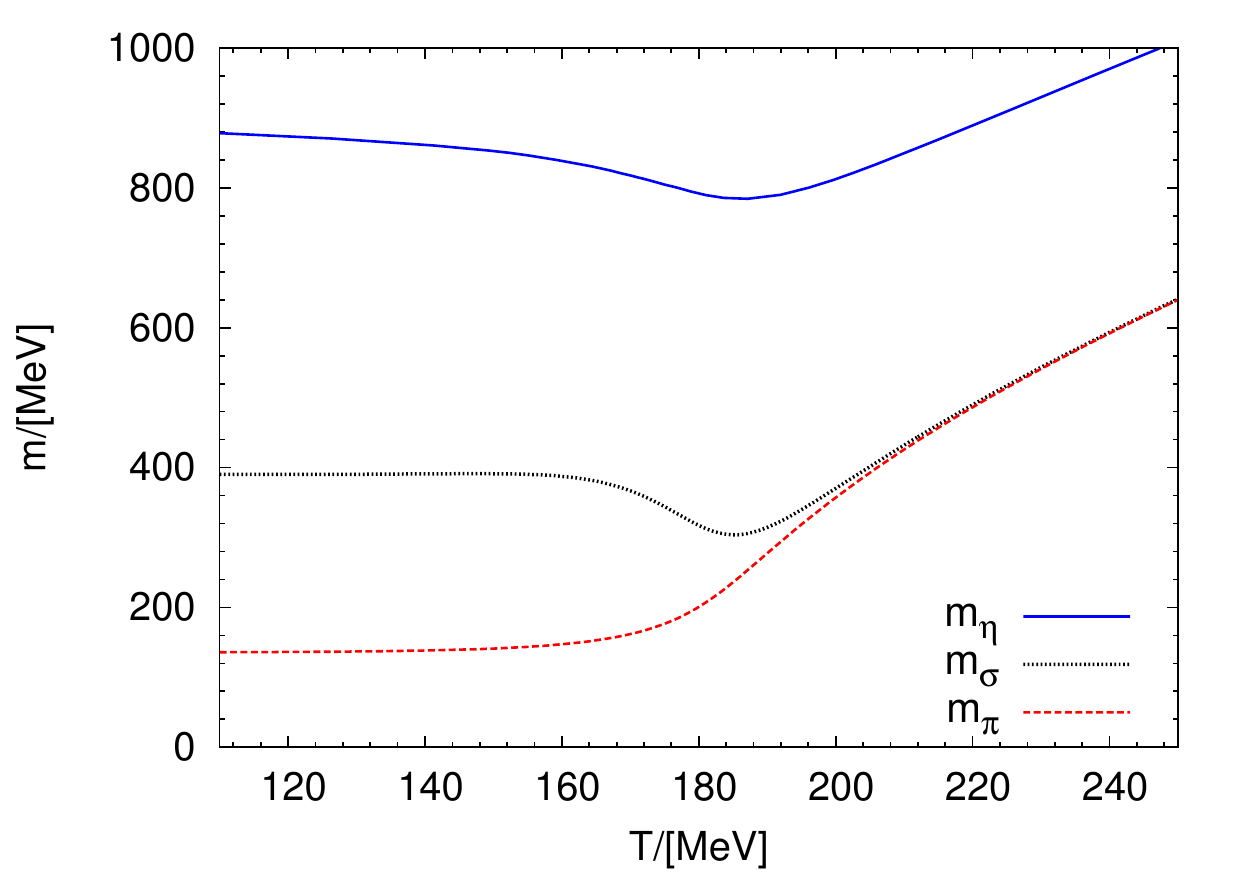}
	    \caption{Meson screening masses as function of the temperature $T$.}
	    \label{fig:mass}      
    \end{figure}

We show the temperature dependence of the mesonic curvature masses, defined by the second derivative of the effective potential, 
in Fig.~\ref{fig:mass}. At small temperatures
our value for the $\eta$-mass is $m_{\eta}= 880$ MeV. This agrees within the given errors with 
lattice QCD which gives approximately $819(127)$ MeV for the corresponding pole mass \cite{Hashimoto:2008xg}.
One uncertainty in our calculation stems from using the curvature mass instead of the pole mass, which is far away from Euclidean momenta. 
Furthermore, we used the assumption that the wave function renormalisation as well as the Yukawa-coupling of the $\eta$-meson 
is degenerate with the one of the pion.

At temperatures close to the crossover we see the usual behaviour of the pion and $\sigma$-meson masses, which become degenerate
above the transition. The mass of the $\eta$-meson shows a drop at the chiral transition.
This is in accordance with experimental results for the in-medium mass \cite{Csorgo:2009pa,Vertesi:2009wf}. 
A similar drop in the $\eta$-meson mass is found in the ($2+1$)-flavour version of the quark-meson model
without temperature-dependence in the 't~Hooft determinant coupling \cite{Schaefer:2008hk,Mitter:2013fxa}.
Since the mesonic 't~Hooft determinant is of order $\Sigma^{N_f}$ in the meson field, the drop in the $(2+1)$-flavour $\eta$-meson mass 
can be attributed solely to the melting of the light condensate. The presented two-flavour case, on the other hand, requires a genuine
temperature dependence in the strength of the 't~Hooft determinant coupling to reproduce a similiar drop. 

Above the chiral crossover, we still see a large splitting between the masses of the pion and the the $\eta$-meson, which is 
in contrast to some lattice simulations reporting a fast reduction of the mass splitting above the chiral crossover \cite{Cossu:2013uua,Cossu:2015lnb}. 
Several assumptions within our calculation can have an influence on the mass-splitting above the chiral crossover. 
Firstly, our procedure of mapping the renormalisation group scale dependence on the temperature introduces uncertainties
precisely at the chiral crossover. Secondly, the restoration of chiral symmetry usually happens at too large temperatures
and too slowly in the quark-meson model.
Consequently, the temperature points above the crossover should actually be compressed in Fig.~\ref{fig:mass}, which would
entail a faster effective restoration of $U(1)_A$. Finally, the asymmetric hadronisation procedure used in \cite{Mitter:2014wpa}
could be responsible for a too large mass splitting in the chirally symmetric phase. 
In future investigations, the $\eta$-channel should therefore be dynamically hadronised as well.

\section{Summary and Conclusion}\label{sec:conclusion}

We have investigated the mass splitting between pions and $\eta$-meson
across the chiral crossover. To this end, we have used results for the 
renormalisation group scale dependence of the 't~Hooft determinant from two-flavour 
quenched QCD. From these results the temperature dependence for the 't~Hooft determinant
has been approximated by matching the temperature
dependence of the pion mass in a mean-field approximation of the Polyakov-quark-meson model to its 
renormalisation group scale dependence from the result in quenched QCD \cite{Mitter:2014wpa}.

We find a drop in the mass of the $\eta$-meson at the chiral crossover which is compatible
with experimental results for the in-medium $\eta$-meson mass \cite{Csorgo:2009pa,Vertesi:2009wf}.
In the case of two flavours, this drop is the consequence of a genuine temperature dependence in
the strength of the 't~Hooft determinant coupling.
A large splitting between pion and $\eta$-meson mass is found at temperatures
above the chiral transition. This might be caused by the slow restoration of chiral symmetry at large
temperatures within the used model together with the procedure used for mapping the
renormalisation group scale to the temperature. Additionally, the asymmetric choice in the 
dynamical hadronisation procedure of \cite{Mitter:2014wpa} and the fact that we used curvature 
masses instead of pole masses might play a r\^ole.

For future investigations it would be interesting to improve the calculation \cite{Mitter:2014wpa}
with respect to axial anomaly effects and dynamically hadronise the $\eta$-meson as well. 
It would be interesting to include mesonic fluctuations in the PQM-model and see whether
the found temperature dependence in the 't~Hooft determinant coupling strength is sufficiently
strong to change the order of the chiral transition in the chiral limit. Furthermore, a continuation 
of the current approach to Minkowski space similar to \cite{Pawlowski:2015mia} would be desirable.

\subsection*{Acknowledgements}
We thank J.~M.~Pawlowski, B.-J.~Schaefer, L.~von~Smekal and N.~Strodthoff for 
discussions and collaborations on related projects. 
This work is supported by 
the FWF through Erwin-Schr\"odinger-Stipendium No. J3507-N27.\\

\bibliographystyle{apsrev4-1}

\begin{thebibliography}{66}%
\makeatletter
\providecommand \@ifxundefined [1]{%
 \@ifx{#1\undefined}
}%
\providecommand \@ifnum [1]{%
 \ifnum #1\expandafter \@firstoftwo
 \else \expandafter \@secondoftwo
 \fi
}%
\providecommand \@ifx [1]{%
 \ifx #1\expandafter \@firstoftwo
 \else \expandafter \@secondoftwo
 \fi
}%
\providecommand \natexlab [1]{#1}%
\providecommand \enquote  [1]{``#1''}%
\providecommand \bibnamefont  [1]{#1}%
\providecommand \bibfnamefont [1]{#1}%
\providecommand \citenamefont [1]{#1}%
\providecommand \href@noop [0]{\@secondoftwo}%
\providecommand \href [0]{\begingroup \@sanitize@url \@href}%
\providecommand \@href[1]{\@@startlink{#1}\@@href}%
\providecommand \@@href[1]{\endgroup#1\@@endlink}%
\providecommand \@sanitize@url [0]{\catcode `\\12\catcode `\$12\catcode
  `\&12\catcode `\#12\catcode `\^12\catcode `\_12\catcode `\%12\relax}%
\providecommand \@@startlink[1]{}%
\providecommand \@@endlink[0]{}%
\providecommand \url  [0]{\begingroup\@sanitize@url \@url }%
\providecommand \@url [1]{\endgroup\@href {#1}{\urlprefix }}%
\providecommand \urlprefix  [0]{URL }%
\providecommand \Eprint [0]{\href }%
\providecommand \doibase [0]{http://dx.doi.org/}%
\providecommand \selectlanguage [0]{\@gobble}%
\providecommand \bibinfo  [0]{\@secondoftwo}%
\providecommand \bibfield  [0]{\@secondoftwo}%
\providecommand \translation [1]{[#1]}%
\providecommand \BibitemOpen [0]{}%
\providecommand \bibitemStop [0]{}%
\providecommand \bibitemNoStop [0]{.\EOS\space}%
\providecommand \EOS [0]{\spacefactor3000\relax}%
\providecommand \BibitemShut  [1]{\csname bibitem#1\endcsname}%
\let\auto@bib@innerbib\@empty
\bibitem [{\citenamefont {Witten}(1979)}]{Witten:1979vv}%
  \BibitemOpen
  \bibfield  {author} {\bibinfo {author} {\bibfnamefont {E.}~\bibnamefont
  {Witten}},\ }\href {\doibase 10.1016/0550-3213(79)90031-2} {\bibfield
  {journal} {\bibinfo  {journal} {Nucl. Phys.}\ }\textbf {\bibinfo {volume}
  {B156}},\ \bibinfo {pages} {269} (\bibinfo {year} {1979})}\BibitemShut
  {NoStop}%
\bibitem [{\citenamefont {Veneziano}(1979)}]{Veneziano:1979ec}%
  \BibitemOpen
  \bibfield  {author} {\bibinfo {author} {\bibfnamefont {G.}~\bibnamefont
  {Veneziano}},\ }\href {\doibase 10.1016/0550-3213(79)90332-8} {\bibfield
  {journal} {\bibinfo  {journal} {Nucl. Phys.}\ }\textbf {\bibinfo {volume}
  {B159}},\ \bibinfo {pages} {213} (\bibinfo {year} {1979})}\BibitemShut
  {NoStop}%
\bibitem [{\citenamefont {'t~Hooft}(1976{\natexlab{a}})}]{'tHooft:1976up}%
  \BibitemOpen
  \bibfield  {author} {\bibinfo {author} {\bibfnamefont {G.}~\bibnamefont
  {'t~Hooft}},\ }\href {\doibase 10.1103/PhysRevLett.37.8} {\bibfield
  {journal} {\bibinfo  {journal} {Phys.Rev.Lett.}\ }\textbf {\bibinfo {volume}
  {37}},\ \bibinfo {pages} {8} (\bibinfo {year}
  {1976}{\natexlab{a}})}\BibitemShut {NoStop}%
\bibitem [{\citenamefont {'t~Hooft}(1976{\natexlab{b}})}]{'tHooft:1976fv}%
  \BibitemOpen
  \bibfield  {author} {\bibinfo {author} {\bibfnamefont {G.}~\bibnamefont
  {'t~Hooft}},\ }\href {\doibase 10.1103/PhysRevD.18.2199.3,
  10.1103/PhysRevD.14.3432} {\bibfield  {journal} {\bibinfo  {journal}
  {Phys.Rev.}\ }\textbf {\bibinfo {volume} {D14}},\ \bibinfo {pages} {3432}
  (\bibinfo {year} {1976}{\natexlab{b}})}\BibitemShut {NoStop}%
\bibitem [{\citenamefont {Mitter}\ \emph {et~al.}(2015)\citenamefont {Mitter},
  \citenamefont {Pawlowski},\ and\ \citenamefont
  {Strodthoff}}]{Mitter:2014wpa}%
  \BibitemOpen
  \bibfield  {author} {\bibinfo {author} {\bibfnamefont {M.}~\bibnamefont
  {Mitter}}, \bibinfo {author} {\bibfnamefont {J.~M.}\ \bibnamefont
  {Pawlowski}}, \ and\ \bibinfo {author} {\bibfnamefont {N.}~\bibnamefont
  {Strodthoff}},\ }\href {\doibase 10.1103/PhysRevD.91.054035} {\bibfield
  {journal} {\bibinfo  {journal} {Phys.Rev.}\ }\textbf {\bibinfo {volume}
  {D91}},\ \bibinfo {pages} {054035} (\bibinfo {year} {2015})},\ \Eprint
  {http://arxiv.org/abs/1411.7978} {arXiv:1411.7978 [hep-ph]} \BibitemShut
  {NoStop}%
\bibitem [{\citenamefont {Pawlowski}(1998)}]{Pawlowski:1996ch}%
  \BibitemOpen
  \bibfield  {author} {\bibinfo {author} {\bibfnamefont {J.}~\bibnamefont
  {Pawlowski}},\ }\href {\doibase 10.1103/PhysRevD.58.045011} {\bibfield
  {journal} {\bibinfo  {journal} {Phys.Rev.}\ }\textbf {\bibinfo {volume}
  {D58}},\ \bibinfo {pages} {045011} (\bibinfo {year} {1998})},\ \Eprint
  {http://arxiv.org/abs/hep-th/9605037} {arXiv:hep-th/9605037 [hep-th]}
  \BibitemShut {NoStop}%
\bibitem [{\citenamefont {Aoki}\ \emph {et~al.}(2006)\citenamefont {Aoki},
  \citenamefont {Endrodi}, \citenamefont {Fodor}, \citenamefont {Katz},\ and\
  \citenamefont {Szabo}}]{Aoki:2006we}%
  \BibitemOpen
  \bibfield  {author} {\bibinfo {author} {\bibfnamefont {Y.}~\bibnamefont
  {Aoki}}, \bibinfo {author} {\bibfnamefont {G.}~\bibnamefont {Endrodi}},
  \bibinfo {author} {\bibfnamefont {Z.}~\bibnamefont {Fodor}}, \bibinfo
  {author} {\bibfnamefont {S.}~\bibnamefont {Katz}}, \ and\ \bibinfo {author}
  {\bibfnamefont {K.}~\bibnamefont {Szabo}},\ }\href {\doibase
  10.1038/nature05120} {\bibfield  {journal} {\bibinfo  {journal} {Nature}\
  }\textbf {\bibinfo {volume} {443}},\ \bibinfo {pages} {675} (\bibinfo {year}
  {2006})},\ \Eprint {http://arxiv.org/abs/hep-lat/0611014}
  {arXiv:hep-lat/0611014 [hep-lat]} \BibitemShut {NoStop}%
\bibitem [{\citenamefont {Aoki}\ \emph {et~al.}(2009)\citenamefont {Aoki},
  \citenamefont {Borsanyi}, \citenamefont {Durr}, \citenamefont {Fodor},
  \citenamefont {Katz} \emph {et~al.}}]{Aoki:2009sc}%
  \BibitemOpen
  \bibfield  {author} {\bibinfo {author} {\bibfnamefont {Y.}~\bibnamefont
  {Aoki}}, \bibinfo {author} {\bibfnamefont {S.}~\bibnamefont {Borsanyi}},
  \bibinfo {author} {\bibfnamefont {S.}~\bibnamefont {Durr}}, \bibinfo {author}
  {\bibfnamefont {Z.}~\bibnamefont {Fodor}}, \bibinfo {author} {\bibfnamefont
  {S.~D.}\ \bibnamefont {Katz}},  \emph {et~al.},\ }\href {\doibase
  10.1088/1126-6708/2009/06/088} {\bibfield  {journal} {\bibinfo  {journal}
  {JHEP}\ }\textbf {\bibinfo {volume} {0906}},\ \bibinfo {pages} {088}
  (\bibinfo {year} {2009})},\ \Eprint {http://arxiv.org/abs/0903.4155}
  {arXiv:0903.4155 [hep-lat]} \BibitemShut {NoStop}%
\bibitem [{\citenamefont {Bhattacharya}\ \emph {et~al.}(2014)\citenamefont
  {Bhattacharya} \emph {et~al.}}]{Bhattacharya:2014ara}%
  \BibitemOpen
  \bibfield  {author} {\bibinfo {author} {\bibfnamefont {T.}~\bibnamefont
  {Bhattacharya}} \emph {et~al.},\ }\href {\doibase
  10.1103/PhysRevLett.113.082001} {\bibfield  {journal} {\bibinfo  {journal}
  {Phys. Rev. Lett.}\ }\textbf {\bibinfo {volume} {113}},\ \bibinfo {pages}
  {082001} (\bibinfo {year} {2014})},\ \Eprint {http://arxiv.org/abs/1402.5175}
  {arXiv:1402.5175 [hep-lat]} \BibitemShut {NoStop}%
\bibitem [{\citenamefont {Buballa}(2005)}]{Buballa:2003qv}%
  \BibitemOpen
  \bibfield  {author} {\bibinfo {author} {\bibfnamefont {M.}~\bibnamefont
  {Buballa}},\ }\href {\doibase 10.1016/j.physrep.2004.11.004} {\bibfield
  {journal} {\bibinfo  {journal} {Phys.Rept.}\ }\textbf {\bibinfo {volume}
  {407}},\ \bibinfo {pages} {205} (\bibinfo {year} {2005})},\ \Eprint
  {http://arxiv.org/abs/hep-ph/0402234} {arXiv:hep-ph/0402234 [hep-ph]}
  \BibitemShut {NoStop}%
\bibitem [{\citenamefont {Schaefer}\ and\ \citenamefont
  {Wambach}(2008)}]{Schaefer:2006sr}%
  \BibitemOpen
  \bibfield  {author} {\bibinfo {author} {\bibfnamefont {B.-J.}\ \bibnamefont
  {Schaefer}}\ and\ \bibinfo {author} {\bibfnamefont {J.}~\bibnamefont
  {Wambach}},\ }\href {\doibase 10.1134/S1063779608070083} {\bibfield
  {journal} {\bibinfo  {journal} {Phys.Part.Nucl.}\ }\textbf {\bibinfo {volume}
  {39}},\ \bibinfo {pages} {1025} (\bibinfo {year} {2008})},\ \Eprint
  {http://arxiv.org/abs/hep-ph/0611191} {arXiv:hep-ph/0611191 [hep-ph]}
  \BibitemShut {NoStop}%
\bibitem [{\citenamefont {Alford}\ \emph {et~al.}(2008)\citenamefont {Alford},
  \citenamefont {Schmitt}, \citenamefont {Rajagopal},\ and\ \citenamefont
  {Sch\"afer}}]{Alford:2007xm}%
  \BibitemOpen
  \bibfield  {author} {\bibinfo {author} {\bibfnamefont {M.~G.}\ \bibnamefont
  {Alford}}, \bibinfo {author} {\bibfnamefont {A.}~\bibnamefont {Schmitt}},
  \bibinfo {author} {\bibfnamefont {K.}~\bibnamefont {Rajagopal}}, \ and\
  \bibinfo {author} {\bibfnamefont {T.}~\bibnamefont {Sch\"afer}},\ }\href
  {\doibase 10.1103/RevModPhys.80.1455} {\bibfield  {journal} {\bibinfo
  {journal} {Rev. Mod. Phys.}\ }\textbf {\bibinfo {volume} {80}},\ \bibinfo
  {pages} {1455} (\bibinfo {year} {2008})},\ \Eprint
  {http://arxiv.org/abs/0709.4635} {arXiv:0709.4635 [hep-ph]} \BibitemShut
  {NoStop}%
\bibitem [{\citenamefont {Fukushima}\ and\ \citenamefont
  {Hatsuda}(2011)}]{Fukushima:2010bq}%
  \BibitemOpen
  \bibfield  {author} {\bibinfo {author} {\bibfnamefont {K.}~\bibnamefont
  {Fukushima}}\ and\ \bibinfo {author} {\bibfnamefont {T.}~\bibnamefont
  {Hatsuda}},\ }\href {\doibase 10.1088/0034-4885/74/1/014001} {\bibfield
  {journal} {\bibinfo  {journal} {Rept.Prog.Phys.}\ }\textbf {\bibinfo {volume}
  {74}},\ \bibinfo {pages} {014001} (\bibinfo {year} {2011})},\ \Eprint
  {http://arxiv.org/abs/1005.4814} {arXiv:1005.4814 [hep-ph]} \BibitemShut
  {NoStop}%
\bibitem [{\citenamefont {Fukushima}\ and\ \citenamefont
  {Sasaki}(2013)}]{Fukushima:2013rx}%
  \BibitemOpen
  \bibfield  {author} {\bibinfo {author} {\bibfnamefont {K.}~\bibnamefont
  {Fukushima}}\ and\ \bibinfo {author} {\bibfnamefont {C.}~\bibnamefont
  {Sasaki}},\ }\href@noop {} {\  (\bibinfo {year} {2013})},\ \Eprint
  {http://arxiv.org/abs/1301.6377} {arXiv:1301.6377 [hep-ph]} \BibitemShut
  {NoStop}%
\bibitem [{\citenamefont {Pawlowski}(2014)}]{Pawlowski:2014aha}%
  \BibitemOpen
  \bibfield  {author} {\bibinfo {author} {\bibfnamefont {J.~M.}\ \bibnamefont
  {Pawlowski}},\ }\href {\doibase 10.1016/j.nuclphysa.2014.09.074} {\bibfield
  {journal} {\bibinfo  {journal} {Nucl.Phys.}\ }\textbf {\bibinfo {volume}
  {A931}},\ \bibinfo {pages} {113} (\bibinfo {year} {2014})}\BibitemShut
  {NoStop}%
\bibitem [{\citenamefont {Gross}\ \emph {et~al.}(1981)\citenamefont {Gross},
  \citenamefont {Pisarski},\ and\ \citenamefont {Yaffe}}]{Gross:1980br}%
  \BibitemOpen
  \bibfield  {author} {\bibinfo {author} {\bibfnamefont {D.~J.}\ \bibnamefont
  {Gross}}, \bibinfo {author} {\bibfnamefont {R.~D.}\ \bibnamefont {Pisarski}},
  \ and\ \bibinfo {author} {\bibfnamefont {L.~G.}\ \bibnamefont {Yaffe}},\
  }\href {\doibase 10.1103/RevModPhys.53.43} {\bibfield  {journal} {\bibinfo
  {journal} {Rev.Mod.Phys.}\ }\textbf {\bibinfo {volume} {53}},\ \bibinfo
  {pages} {43} (\bibinfo {year} {1981})}\BibitemShut {NoStop}%
\bibitem [{\citenamefont {Sch\"afer}\ and\ \citenamefont
  {Shuryak}(1998)}]{Schafer:1996wv}%
  \BibitemOpen
  \bibfield  {author} {\bibinfo {author} {\bibfnamefont {T.}~\bibnamefont
  {Sch\"afer}}\ and\ \bibinfo {author} {\bibfnamefont {E.~V.}\ \bibnamefont
  {Shuryak}},\ }\href {\doibase 10.1103/RevModPhys.70.323} {\bibfield
  {journal} {\bibinfo  {journal} {Rev. Mod. Phys.}\ }\textbf {\bibinfo {volume}
  {70}},\ \bibinfo {pages} {323} (\bibinfo {year} {1998})},\ \Eprint
  {http://arxiv.org/abs/hep-ph/9610451} {arXiv:hep-ph/9610451 [hep-ph]}
  \BibitemShut {NoStop}%
\bibitem [{\citenamefont {Kapusta}\ \emph {et~al.}(1996)\citenamefont
  {Kapusta}, \citenamefont {Kharzeev},\ and\ \citenamefont
  {McLerran}}]{Kapusta:1995ww}%
  \BibitemOpen
  \bibfield  {author} {\bibinfo {author} {\bibfnamefont {J.~I.}\ \bibnamefont
  {Kapusta}}, \bibinfo {author} {\bibfnamefont {D.}~\bibnamefont {Kharzeev}}, \
  and\ \bibinfo {author} {\bibfnamefont {L.~D.}\ \bibnamefont {McLerran}},\
  }\href {\doibase 10.1103/PhysRevD.53.5028} {\bibfield  {journal} {\bibinfo
  {journal} {Phys. Rev.}\ }\textbf {\bibinfo {volume} {D53}},\ \bibinfo {pages}
  {5028} (\bibinfo {year} {1996})},\ \Eprint
  {http://arxiv.org/abs/hep-ph/9507343} {arXiv:hep-ph/9507343 [hep-ph]}
  \BibitemShut {NoStop}%
\bibitem [{\citenamefont {Csorgo}\ \emph {et~al.}(2010)\citenamefont {Csorgo},
  \citenamefont {Vertesi},\ and\ \citenamefont {Sziklai}}]{Csorgo:2009pa}%
  \BibitemOpen
  \bibfield  {author} {\bibinfo {author} {\bibfnamefont {T.}~\bibnamefont
  {Csorgo}}, \bibinfo {author} {\bibfnamefont {R.}~\bibnamefont {Vertesi}}, \
  and\ \bibinfo {author} {\bibfnamefont {J.}~\bibnamefont {Sziklai}},\ }\href
  {\doibase 10.1103/PhysRevLett.105.182301} {\bibfield  {journal} {\bibinfo
  {journal} {Phys. Rev. Lett.}\ }\textbf {\bibinfo {volume} {105}},\ \bibinfo
  {pages} {182301} (\bibinfo {year} {2010})},\ \Eprint
  {http://arxiv.org/abs/0912.5526} {arXiv:0912.5526 [nucl-ex]} \BibitemShut
  {NoStop}%
\bibitem [{\citenamefont {Vertesi}\ \emph {et~al.}(2011)\citenamefont
  {Vertesi}, \citenamefont {Csorgo},\ and\ \citenamefont
  {Sziklai}}]{Vertesi:2009wf}%
  \BibitemOpen
  \bibfield  {author} {\bibinfo {author} {\bibfnamefont {R.}~\bibnamefont
  {Vertesi}}, \bibinfo {author} {\bibfnamefont {T.}~\bibnamefont {Csorgo}}, \
  and\ \bibinfo {author} {\bibfnamefont {J.}~\bibnamefont {Sziklai}},\ }\href
  {\doibase 10.1103/PhysRevC.83.054903} {\bibfield  {journal} {\bibinfo
  {journal} {Phys. Rev.}\ }\textbf {\bibinfo {volume} {C83}},\ \bibinfo {pages}
  {054903} (\bibinfo {year} {2011})},\ \Eprint {http://arxiv.org/abs/0912.0258}
  {arXiv:0912.0258 [nucl-ex]} \BibitemShut {NoStop}%
\bibitem [{\citenamefont {Pisarski}\ and\ \citenamefont
  {Wilczek}(1984)}]{Pisarski:1983ms}%
  \BibitemOpen
  \bibfield  {author} {\bibinfo {author} {\bibfnamefont {R.~D.}\ \bibnamefont
  {Pisarski}}\ and\ \bibinfo {author} {\bibfnamefont {F.}~\bibnamefont
  {Wilczek}},\ }\href {\doibase 10.1103/PhysRevD.29.338} {\bibfield  {journal}
  {\bibinfo  {journal} {Phys.Rev.}\ }\textbf {\bibinfo {volume} {D29}},\
  \bibinfo {pages} {338} (\bibinfo {year} {1984})}\BibitemShut {NoStop}%
\bibitem [{\citenamefont {Butti}\ \emph {et~al.}(2003)\citenamefont {Butti},
  \citenamefont {Pelissetto},\ and\ \citenamefont {Vicari}}]{Butti:2003nu}%
  \BibitemOpen
  \bibfield  {author} {\bibinfo {author} {\bibfnamefont {A.}~\bibnamefont
  {Butti}}, \bibinfo {author} {\bibfnamefont {A.}~\bibnamefont {Pelissetto}}, \
  and\ \bibinfo {author} {\bibfnamefont {E.}~\bibnamefont {Vicari}},\ }\href
  {\doibase 10.1088/1126-6708/2003/08/029} {\bibfield  {journal} {\bibinfo
  {journal} {JHEP}\ }\textbf {\bibinfo {volume} {08}},\ \bibinfo {pages} {029}
  (\bibinfo {year} {2003})},\ \Eprint {http://arxiv.org/abs/hep-ph/0307036}
  {arXiv:hep-ph/0307036 [hep-ph]} \BibitemShut {NoStop}%
\bibitem [{\citenamefont {Pelissetto}\ and\ \citenamefont
  {Vicari}(2013)}]{Pelissetto:2013hqa}%
  \BibitemOpen
  \bibfield  {author} {\bibinfo {author} {\bibfnamefont {A.}~\bibnamefont
  {Pelissetto}}\ and\ \bibinfo {author} {\bibfnamefont {E.}~\bibnamefont
  {Vicari}},\ }\href {\doibase 10.1103/PhysRevD.88.105018} {\bibfield
  {journal} {\bibinfo  {journal} {Phys. Rev.}\ }\textbf {\bibinfo {volume}
  {D88}},\ \bibinfo {pages} {105018} (\bibinfo {year} {2013})},\ \Eprint
  {http://arxiv.org/abs/1309.5446} {arXiv:1309.5446 [hep-lat]} \BibitemShut
  {NoStop}%
\bibitem [{\citenamefont {Meggiolaro}\ and\ \citenamefont
  {Morda}(2013)}]{Meggiolaro:2013swa}%
  \BibitemOpen
  \bibfield  {author} {\bibinfo {author} {\bibfnamefont {E.}~\bibnamefont
  {Meggiolaro}}\ and\ \bibinfo {author} {\bibfnamefont {A.}~\bibnamefont
  {Morda}},\ }\href {\doibase 10.1103/PhysRevD.88.096010} {\bibfield  {journal}
  {\bibinfo  {journal} {Phys. Rev.}\ }\textbf {\bibinfo {volume} {D88}},\
  \bibinfo {pages} {096010} (\bibinfo {year} {2013})},\ \Eprint
  {http://arxiv.org/abs/1309.4598} {arXiv:1309.4598 [hep-ph]} \BibitemShut
  {NoStop}%
\bibitem [{\citenamefont {Grahl}\ and\ \citenamefont
  {Rischke}(2013)}]{Grahl:2013pba}%
  \BibitemOpen
  \bibfield  {author} {\bibinfo {author} {\bibfnamefont {M.}~\bibnamefont
  {Grahl}}\ and\ \bibinfo {author} {\bibfnamefont {D.~H.}\ \bibnamefont
  {Rischke}},\ }\href {\doibase 10.1103/PhysRevD.88.056014} {\bibfield
  {journal} {\bibinfo  {journal} {Phys. Rev.}\ }\textbf {\bibinfo {volume}
  {D88}},\ \bibinfo {pages} {056014} (\bibinfo {year} {2013})},\ \Eprint
  {http://arxiv.org/abs/1307.2184} {arXiv:1307.2184 [hep-th]} \BibitemShut
  {NoStop}%
\bibitem [{\citenamefont {Grahl}(2014)}]{Grahl:2014fna}%
  \BibitemOpen
  \bibfield  {author} {\bibinfo {author} {\bibfnamefont {M.}~\bibnamefont
  {Grahl}},\ }\href {\doibase 10.1103/PhysRevD.90.117904} {\bibfield  {journal}
  {\bibinfo  {journal} {Phys. Rev.}\ }\textbf {\bibinfo {volume} {D90}},\
  \bibinfo {pages} {117904} (\bibinfo {year} {2014})},\ \Eprint
  {http://arxiv.org/abs/1410.0985} {arXiv:1410.0985 [hep-th]} \BibitemShut
  {NoStop}%
\bibitem [{\citenamefont {Cossu}\ \emph {et~al.}(2013)\citenamefont {Cossu},
  \citenamefont {Aoki}, \citenamefont {Fukaya}, \citenamefont {Hashimoto},
  \citenamefont {Kaneko}, \citenamefont {Matsufuru},\ and\ \citenamefont
  {Noaki}}]{Cossu:2013uua}%
  \BibitemOpen
  \bibfield  {author} {\bibinfo {author} {\bibfnamefont {G.}~\bibnamefont
  {Cossu}}, \bibinfo {author} {\bibfnamefont {S.}~\bibnamefont {Aoki}},
  \bibinfo {author} {\bibfnamefont {H.}~\bibnamefont {Fukaya}}, \bibinfo
  {author} {\bibfnamefont {S.}~\bibnamefont {Hashimoto}}, \bibinfo {author}
  {\bibfnamefont {T.}~\bibnamefont {Kaneko}}, \bibinfo {author} {\bibfnamefont
  {H.}~\bibnamefont {Matsufuru}}, \ and\ \bibinfo {author} {\bibfnamefont
  {J.-I.}\ \bibnamefont {Noaki}},\ }\href {\doibase 10.1103/PhysRevD.88.019901,
  10.1103/PhysRevD.87.114514} {\bibfield  {journal} {\bibinfo  {journal} {Phys.
  Rev.}\ }\textbf {\bibinfo {volume} {D87}},\ \bibinfo {pages} {114514}
  (\bibinfo {year} {2013})},\ \bibinfo {note} {[Erratum: Phys.
  Rev.D88,no.1,019901(2013)]},\ \Eprint {http://arxiv.org/abs/1304.6145}
  {arXiv:1304.6145 [hep-lat]} \BibitemShut {NoStop}%
\bibitem [{\citenamefont {Cossu}\ \emph {et~al.}(2015)\citenamefont {Cossu},
  \citenamefont {Fukaya}, \citenamefont {Hashimoto}, \citenamefont {Noaki},\
  and\ \citenamefont {Tomiya}}]{Cossu:2015lnb}%
  \BibitemOpen
  \bibfield  {author} {\bibinfo {author} {\bibfnamefont {G.}~\bibnamefont
  {Cossu}}, \bibinfo {author} {\bibfnamefont {H.}~\bibnamefont {Fukaya}},
  \bibinfo {author} {\bibfnamefont {S.}~\bibnamefont {Hashimoto}}, \bibinfo
  {author} {\bibfnamefont {J.-i.}\ \bibnamefont {Noaki}}, \ and\ \bibinfo
  {author} {\bibfnamefont {A.}~\bibnamefont {Tomiya}} (\bibinfo {collaboration}
  {JLQCD}),\ }in\ \href
  {http://inspirehep.net/record/1405303/files/arXiv:1511.05691.pdf} {\emph
  {\bibinfo {booktitle} {{Proceedings, 33rd International Symposium on Lattice
  Field Theory (Lattice 2015)}}}}\ (\bibinfo {year} {2015})\ \Eprint
  {http://arxiv.org/abs/1511.05691} {arXiv:1511.05691 [hep-lat]} \BibitemShut
  {NoStop}%
\bibitem [{\citenamefont {Bazavov}\ \emph {et~al.}(2012)\citenamefont {Bazavov}
  \emph {et~al.}}]{Bazavov:2012qja}%
  \BibitemOpen
  \bibfield  {author} {\bibinfo {author} {\bibfnamefont {A.}~\bibnamefont
  {Bazavov}} \emph {et~al.} (\bibinfo {collaboration} {HotQCD}),\ }\href
  {\doibase 10.1103/PhysRevD.86.094503} {\bibfield  {journal} {\bibinfo
  {journal} {Phys. Rev.}\ }\textbf {\bibinfo {volume} {D86}},\ \bibinfo {pages}
  {094503} (\bibinfo {year} {2012})},\ \Eprint {http://arxiv.org/abs/1205.3535}
  {arXiv:1205.3535 [hep-lat]} \BibitemShut {NoStop}%
\bibitem [{\citenamefont {Sharma}\ \emph {et~al.}(2014)\citenamefont {Sharma},
  \citenamefont {Dick}, \citenamefont {Karsch}, \citenamefont {Laermann},\ and\
  \citenamefont {Mukherjee}}]{Sharma:2013nva}%
  \BibitemOpen
  \bibfield  {author} {\bibinfo {author} {\bibfnamefont {S.}~\bibnamefont
  {Sharma}}, \bibinfo {author} {\bibfnamefont {V.}~\bibnamefont {Dick}},
  \bibinfo {author} {\bibfnamefont {F.}~\bibnamefont {Karsch}}, \bibinfo
  {author} {\bibfnamefont {E.}~\bibnamefont {Laermann}}, \ and\ \bibinfo
  {author} {\bibfnamefont {S.}~\bibnamefont {Mukherjee}},\ }\bibfield
  {booktitle} {\emph {\bibinfo {booktitle} {{Proceedings, 31st International
  Symposium on Lattice Field Theory (Lattice 2013)}}},\ }\href@noop {}
  {\bibfield  {journal} {\bibinfo  {journal} {PoS}\ }\textbf {\bibinfo {volume}
  {LATTICE2013}},\ \bibinfo {pages} {164} (\bibinfo {year} {2014})},\ \Eprint
  {http://arxiv.org/abs/1311.3943} {arXiv:1311.3943 [hep-lat]} \BibitemShut
  {NoStop}%
\bibitem [{\citenamefont {Dick}\ \emph {et~al.}(2015)\citenamefont {Dick},
  \citenamefont {Karsch}, \citenamefont {Laermann}, \citenamefont {Mukherjee},\
  and\ \citenamefont {Sharma}}]{Dick:2015twa}%
  \BibitemOpen
  \bibfield  {author} {\bibinfo {author} {\bibfnamefont {V.}~\bibnamefont
  {Dick}}, \bibinfo {author} {\bibfnamefont {F.}~\bibnamefont {Karsch}},
  \bibinfo {author} {\bibfnamefont {E.}~\bibnamefont {Laermann}}, \bibinfo
  {author} {\bibfnamefont {S.}~\bibnamefont {Mukherjee}}, \ and\ \bibinfo
  {author} {\bibfnamefont {S.}~\bibnamefont {Sharma}},\ }\href {\doibase
  10.1103/PhysRevD.91.094504} {\bibfield  {journal} {\bibinfo  {journal} {Phys.
  Rev.}\ }\textbf {\bibinfo {volume} {D91}},\ \bibinfo {pages} {094504}
  (\bibinfo {year} {2015})},\ \Eprint {http://arxiv.org/abs/1502.06190}
  {arXiv:1502.06190 [hep-lat]} \BibitemShut {NoStop}%
\bibitem [{\citenamefont {von Smekal}\ \emph {et~al.}(1997)\citenamefont {von
  Smekal}, \citenamefont {Mecke},\ and\ \citenamefont
  {Alkofer}}]{vonSmekal:1997dq}%
  \BibitemOpen
  \bibfield  {author} {\bibinfo {author} {\bibfnamefont {L.}~\bibnamefont {von
  Smekal}}, \bibinfo {author} {\bibfnamefont {A.}~\bibnamefont {Mecke}}, \ and\
  \bibinfo {author} {\bibfnamefont {R.}~\bibnamefont {Alkofer}},\ }\bibfield
  {booktitle} {\emph {\bibinfo {booktitle} {{In *Big Sky 1997, Intersections
  between particle and nuclear physics* 746-749}}},\ }\href {\doibase
  10.1063/1.54300} {\  (\bibinfo {year} {1997}),\ 10.1063/1.54300},\ \bibinfo
  {note} {[AIP Conf. Proc.412,746(1997)]},\ \Eprint
  {http://arxiv.org/abs/hep-ph/9707210} {arXiv:hep-ph/9707210 [hep-ph]}
  \BibitemShut {NoStop}%
\bibitem [{\citenamefont {Alkofer}\ \emph {et~al.}(2008)\citenamefont
  {Alkofer}, \citenamefont {Fischer},\ and\ \citenamefont
  {Williams}}]{Alkofer:2008et}%
  \BibitemOpen
  \bibfield  {author} {\bibinfo {author} {\bibfnamefont {R.}~\bibnamefont
  {Alkofer}}, \bibinfo {author} {\bibfnamefont {C.~S.}\ \bibnamefont
  {Fischer}}, \ and\ \bibinfo {author} {\bibfnamefont {R.}~\bibnamefont
  {Williams}},\ }\href {\doibase 10.1140/epja/i2008-10646-x} {\bibfield
  {journal} {\bibinfo  {journal} {Eur. Phys. J.}\ }\textbf {\bibinfo {volume}
  {A38}},\ \bibinfo {pages} {53} (\bibinfo {year} {2008})},\ \Eprint
  {http://arxiv.org/abs/0804.3478} {arXiv:0804.3478 [hep-ph]} \BibitemShut
  {NoStop}%
\bibitem [{\citenamefont {Benic}\ \emph {et~al.}(2011)\citenamefont {Benic},
  \citenamefont {Horvatic}, \citenamefont {Kekez},\ and\ \citenamefont
  {Klabucar}}]{Benic:2011fv}%
  \BibitemOpen
  \bibfield  {author} {\bibinfo {author} {\bibfnamefont {S.}~\bibnamefont
  {Benic}}, \bibinfo {author} {\bibfnamefont {D.}~\bibnamefont {Horvatic}},
  \bibinfo {author} {\bibfnamefont {D.}~\bibnamefont {Kekez}}, \ and\ \bibinfo
  {author} {\bibfnamefont {D.}~\bibnamefont {Klabucar}},\ }\href {\doibase
  10.1103/PhysRevD.84.016006} {\bibfield  {journal} {\bibinfo  {journal} {Phys.
  Rev.}\ }\textbf {\bibinfo {volume} {D84}},\ \bibinfo {pages} {016006}
  (\bibinfo {year} {2011})},\ \Eprint {http://arxiv.org/abs/1105.0356}
  {arXiv:1105.0356 [hep-ph]} \BibitemShut {NoStop}%
\bibitem [{\citenamefont {Benic}\ \emph {et~al.}(2014)\citenamefont {Benic},
  \citenamefont {Horvatic}, \citenamefont {Kekez},\ and\ \citenamefont
  {Klabucar}}]{Benic:2014mha}%
  \BibitemOpen
  \bibfield  {author} {\bibinfo {author} {\bibfnamefont {S.}~\bibnamefont
  {Benic}}, \bibinfo {author} {\bibfnamefont {D.}~\bibnamefont {Horvatic}},
  \bibinfo {author} {\bibfnamefont {D.}~\bibnamefont {Kekez}}, \ and\ \bibinfo
  {author} {\bibfnamefont {D.}~\bibnamefont {Klabucar}},\ }\href {\doibase
  10.1016/j.physletb.2014.09.029} {\bibfield  {journal} {\bibinfo  {journal}
  {Phys. Lett.}\ }\textbf {\bibinfo {volume} {B738}},\ \bibinfo {pages} {113}
  (\bibinfo {year} {2014})},\ \Eprint {http://arxiv.org/abs/1405.3299}
  {arXiv:1405.3299 [hep-ph]} \BibitemShut {NoStop}%
\bibitem [{\citenamefont {Chen}\ \emph {et~al.}(2009)\citenamefont {Chen},
  \citenamefont {Fukushima}, \citenamefont {Kohyama}, \citenamefont {Ohnishi},\
  and\ \citenamefont {Raha}}]{Chen:2009gv}%
  \BibitemOpen
  \bibfield  {author} {\bibinfo {author} {\bibfnamefont {J.-W.}\ \bibnamefont
  {Chen}}, \bibinfo {author} {\bibfnamefont {K.}~\bibnamefont {Fukushima}},
  \bibinfo {author} {\bibfnamefont {H.}~\bibnamefont {Kohyama}}, \bibinfo
  {author} {\bibfnamefont {K.}~\bibnamefont {Ohnishi}}, \ and\ \bibinfo
  {author} {\bibfnamefont {U.}~\bibnamefont {Raha}},\ }\href {\doibase
  10.1103/PhysRevD.80.054012} {\bibfield  {journal} {\bibinfo  {journal} {Phys.
  Rev.}\ }\textbf {\bibinfo {volume} {D80}},\ \bibinfo {pages} {054012}
  (\bibinfo {year} {2009})},\ \Eprint {http://arxiv.org/abs/0901.2407}
  {arXiv:0901.2407 [hep-ph]} \BibitemShut {NoStop}%
\bibitem [{\citenamefont {Zhang}\ and\ \citenamefont
  {Kunihiro}(2011)}]{Zhang:2011xi}%
  \BibitemOpen
  \bibfield  {author} {\bibinfo {author} {\bibfnamefont {Z.}~\bibnamefont
  {Zhang}}\ and\ \bibinfo {author} {\bibfnamefont {T.}~\bibnamefont
  {Kunihiro}},\ }\href {\doibase 10.1103/PhysRevD.83.114003} {\bibfield
  {journal} {\bibinfo  {journal} {Phys. Rev.}\ }\textbf {\bibinfo {volume}
  {D83}},\ \bibinfo {pages} {114003} (\bibinfo {year} {2011})},\ \Eprint
  {http://arxiv.org/abs/1102.3263} {arXiv:1102.3263 [hep-ph]} \BibitemShut
  {NoStop}%
\bibitem [{\citenamefont {Powell}\ and\ \citenamefont
  {Baym}(2012)}]{Powell:2011ig}%
  \BibitemOpen
  \bibfield  {author} {\bibinfo {author} {\bibfnamefont {P.~D.}\ \bibnamefont
  {Powell}}\ and\ \bibinfo {author} {\bibfnamefont {G.}~\bibnamefont {Baym}},\
  }\href {\doibase 10.1103/PhysRevD.85.074003} {\bibfield  {journal} {\bibinfo
  {journal} {Phys. Rev.}\ }\textbf {\bibinfo {volume} {D85}},\ \bibinfo {pages}
  {074003} (\bibinfo {year} {2012})},\ \Eprint {http://arxiv.org/abs/1111.5911}
  {arXiv:1111.5911 [hep-ph]} \BibitemShut {NoStop}%
\bibitem [{\citenamefont {Bratovic}\ \emph {et~al.}(2013)\citenamefont
  {Bratovic}, \citenamefont {Hatsuda},\ and\ \citenamefont
  {Weise}}]{Bratovic:2012qs}%
  \BibitemOpen
  \bibfield  {author} {\bibinfo {author} {\bibfnamefont {N.~M.}\ \bibnamefont
  {Bratovic}}, \bibinfo {author} {\bibfnamefont {T.}~\bibnamefont {Hatsuda}}, \
  and\ \bibinfo {author} {\bibfnamefont {W.}~\bibnamefont {Weise}},\ }\href
  {\doibase 10.1016/j.physletb.2013.01.003} {\bibfield  {journal} {\bibinfo
  {journal} {Phys.Lett.}\ }\textbf {\bibinfo {volume} {B719}},\ \bibinfo
  {pages} {131} (\bibinfo {year} {2013})},\ \Eprint
  {http://arxiv.org/abs/1204.3788} {arXiv:1204.3788 [hep-ph]} \BibitemShut
  {NoStop}%
\bibitem [{\citenamefont {Schaefer}\ and\ \citenamefont
  {Wagner}(2009)}]{Schaefer:2008hk}%
  \BibitemOpen
  \bibfield  {author} {\bibinfo {author} {\bibfnamefont {B.-J.}\ \bibnamefont
  {Schaefer}}\ and\ \bibinfo {author} {\bibfnamefont {M.}~\bibnamefont
  {Wagner}},\ }\href {\doibase 10.1103/PhysRevD.79.014018} {\bibfield
  {journal} {\bibinfo  {journal} {Phys.Rev.}\ }\textbf {\bibinfo {volume}
  {D79}},\ \bibinfo {pages} {014018} (\bibinfo {year} {2009})},\ \Eprint
  {http://arxiv.org/abs/0808.1491} {arXiv:0808.1491 [hep-ph]} \BibitemShut
  {NoStop}%
\bibitem [{\citenamefont {Mitter}\ and\ \citenamefont
  {Schaefer}(2014)}]{Mitter:2013fxa}%
  \BibitemOpen
  \bibfield  {author} {\bibinfo {author} {\bibfnamefont {M.}~\bibnamefont
  {Mitter}}\ and\ \bibinfo {author} {\bibfnamefont {B.-J.}\ \bibnamefont
  {Schaefer}},\ }\href {\doibase 10.1103/PhysRevD.89.054027} {\bibfield
  {journal} {\bibinfo  {journal} {Phys.Rev.}\ }\textbf {\bibinfo {volume}
  {D89}},\ \bibinfo {pages} {054027} (\bibinfo {year} {2014})},\ \Eprint
  {http://arxiv.org/abs/1308.3176} {arXiv:1308.3176 [hep-ph]} \BibitemShut
  {NoStop}%
\bibitem [{\citenamefont {Herbst}\ \emph {et~al.}(2014)\citenamefont {Herbst},
  \citenamefont {Mitter}, \citenamefont {Pawlowski}, \citenamefont {Schaefer},\
  and\ \citenamefont {Stiele}}]{Herbst:2013ufa}%
  \BibitemOpen
  \bibfield  {author} {\bibinfo {author} {\bibfnamefont {T.~K.}\ \bibnamefont
  {Herbst}}, \bibinfo {author} {\bibfnamefont {M.}~\bibnamefont {Mitter}},
  \bibinfo {author} {\bibfnamefont {J.~M.}\ \bibnamefont {Pawlowski}}, \bibinfo
  {author} {\bibfnamefont {B.-J.}\ \bibnamefont {Schaefer}}, \ and\ \bibinfo
  {author} {\bibfnamefont {R.}~\bibnamefont {Stiele}},\ }\href {\doibase
  10.1016/j.physletb.2014.02.045} {\bibfield  {journal} {\bibinfo  {journal}
  {Phys.Lett.}\ }\textbf {\bibinfo {volume} {B731}},\ \bibinfo {pages} {248}
  (\bibinfo {year} {2014})},\ \Eprint {http://arxiv.org/abs/1308.3621}
  {arXiv:1308.3621 [hep-ph]} \BibitemShut {NoStop}%
\bibitem [{\citenamefont {Fejos}(2015)}]{Fejos:2015xca}%
  \BibitemOpen
  \bibfield  {author} {\bibinfo {author} {\bibfnamefont {G.}~\bibnamefont
  {Fejos}},\ }\href@noop {} {\  (\bibinfo {year} {2015})},\ \Eprint
  {http://arxiv.org/abs/1506.07399} {arXiv:1506.07399 [hep-ph]} \BibitemShut
  {NoStop}%
\bibitem [{\citenamefont {Eser}\ \emph {et~al.}(2015)\citenamefont {Eser},
  \citenamefont {Grahl},\ and\ \citenamefont {Rischke}}]{Eser:2015pka}%
  \BibitemOpen
  \bibfield  {author} {\bibinfo {author} {\bibfnamefont {J.}~\bibnamefont
  {Eser}}, \bibinfo {author} {\bibfnamefont {M.}~\bibnamefont {Grahl}}, \ and\
  \bibinfo {author} {\bibfnamefont {D.~H.}\ \bibnamefont {Rischke}},\
  }\href@noop {} {\  (\bibinfo {year} {2015})},\ \Eprint
  {http://arxiv.org/abs/1508.06928} {arXiv:1508.06928 [hep-ph]} \BibitemShut
  {NoStop}%
\bibitem [{\citenamefont {Wetterich}(1993)}]{Wetterich:1992yh}%
  \BibitemOpen
  \bibfield  {author} {\bibinfo {author} {\bibfnamefont {C.}~\bibnamefont
  {Wetterich}},\ }\href {\doibase 10.1016/0370-2693(93)90726-X} {\bibfield
  {journal} {\bibinfo  {journal} {Phys.Lett.}\ }\textbf {\bibinfo {volume}
  {B301}},\ \bibinfo {pages} {90} (\bibinfo {year} {1993})}\BibitemShut
  {NoStop}%
\bibitem [{\citenamefont {Berges}\ \emph {et~al.}(2002)\citenamefont {Berges},
  \citenamefont {Tetradis},\ and\ \citenamefont {Wetterich}}]{Berges:2000ew}%
  \BibitemOpen
  \bibfield  {author} {\bibinfo {author} {\bibfnamefont {J.}~\bibnamefont
  {Berges}}, \bibinfo {author} {\bibfnamefont {N.}~\bibnamefont {Tetradis}}, \
  and\ \bibinfo {author} {\bibfnamefont {C.}~\bibnamefont {Wetterich}},\ }\href
  {\doibase 10.1016/S0370-1573(01)00098-9} {\bibfield  {journal} {\bibinfo
  {journal} {Phys. Rept.}\ }\textbf {\bibinfo {volume} {363}},\ \bibinfo
  {pages} {223} (\bibinfo {year} {2002})},\ \Eprint
  {http://arxiv.org/abs/hep-ph/0005122} {arXiv:hep-ph/0005122} \BibitemShut
  {NoStop}%
\bibitem [{\citenamefont {Roberts}\ and\ \citenamefont
  {Schmidt}(2000)}]{Roberts:2000aa}%
  \BibitemOpen
  \bibfield  {author} {\bibinfo {author} {\bibfnamefont {C.~D.}\ \bibnamefont
  {Roberts}}\ and\ \bibinfo {author} {\bibfnamefont {S.~M.}\ \bibnamefont
  {Schmidt}},\ }\href@noop {} {\bibfield  {journal} {\bibinfo  {journal} {Prog.
  Part. Nucl. Phys.}\ }\textbf {\bibinfo {volume} {45}},\ \bibinfo {pages} {S1}
  (\bibinfo {year} {2000})},\ \Eprint {http://arxiv.org/abs/nucl-th/0005064}
  {arXiv:nucl-th/0005064} \BibitemShut {NoStop}%
\bibitem [{\citenamefont {Alkofer}\ and\ \citenamefont {von
  Smekal}(2001)}]{Alkofer:2000wg}%
  \BibitemOpen
  \bibfield  {author} {\bibinfo {author} {\bibfnamefont {R.}~\bibnamefont
  {Alkofer}}\ and\ \bibinfo {author} {\bibfnamefont {L.}~\bibnamefont {von
  Smekal}},\ }\href {\doibase 10.1016/S0370-1573(01)00010-2} {\bibfield
  {journal} {\bibinfo  {journal} {Phys. Rept.}\ }\textbf {\bibinfo {volume}
  {353}},\ \bibinfo {pages} {281} (\bibinfo {year} {2001})},\ \Eprint
  {http://arxiv.org/abs/hep-ph/0007355} {arXiv:hep-ph/0007355} \BibitemShut
  {NoStop}%
\bibitem [{\citenamefont {Pawlowski}(2007)}]{Pawlowski:2005xe}%
  \BibitemOpen
  \bibfield  {author} {\bibinfo {author} {\bibfnamefont {J.~M.}\ \bibnamefont
  {Pawlowski}},\ }\href {\doibase 10.1016/j.aop.2007.01.007} {\bibfield
  {journal} {\bibinfo  {journal} {Annals Phys.}\ }\textbf {\bibinfo {volume}
  {322}},\ \bibinfo {pages} {2831} (\bibinfo {year} {2007})},\ \Eprint
  {http://arxiv.org/abs/hep-th/0512261} {arXiv:hep-th/0512261 [hep-th]}
  \BibitemShut {NoStop}%
\bibitem [{\citenamefont {Fischer}(2006)}]{Fischer:2006ub}%
  \BibitemOpen
  \bibfield  {author} {\bibinfo {author} {\bibfnamefont {C.~S.}\ \bibnamefont
  {Fischer}},\ }\href {\doibase 10.1088/0954-3899/32/8/R02} {\bibfield
  {journal} {\bibinfo  {journal} {J.Phys.G}\ }\textbf {\bibinfo {volume}
  {G32}},\ \bibinfo {pages} {R253} (\bibinfo {year} {2006})},\ \Eprint
  {http://arxiv.org/abs/hep-ph/0605173} {arXiv:hep-ph/0605173 [hep-ph]}
  \BibitemShut {NoStop}%
\bibitem [{\citenamefont {Gies}(2012)}]{Gies:2006wv}%
  \BibitemOpen
  \bibfield  {author} {\bibinfo {author} {\bibfnamefont {H.}~\bibnamefont
  {Gies}},\ }\href {\doibase 10.1007/978-3-642-27320-9_6} {\bibfield  {journal}
  {\bibinfo  {journal} {Lect.Notes Phys.}\ }\textbf {\bibinfo {volume} {852}},\
  \bibinfo {pages} {287} (\bibinfo {year} {2012})},\ \Eprint
  {http://arxiv.org/abs/hep-ph/0611146} {arXiv:hep-ph/0611146 [hep-ph]}
  \BibitemShut {NoStop}%
\bibitem [{\citenamefont {Binosi}\ and\ \citenamefont
  {Papavassiliou}(2009)}]{Binosi:2009qm}%
  \BibitemOpen
  \bibfield  {author} {\bibinfo {author} {\bibfnamefont {D.}~\bibnamefont
  {Binosi}}\ and\ \bibinfo {author} {\bibfnamefont {J.}~\bibnamefont
  {Papavassiliou}},\ }\href {\doibase 10.1016/j.physrep.2009.05.001} {\bibfield
   {journal} {\bibinfo  {journal} {Phys.Rept.}\ }\textbf {\bibinfo {volume}
  {479}},\ \bibinfo {pages} {1} (\bibinfo {year} {2009})},\ \Eprint
  {http://arxiv.org/abs/0909.2536} {arXiv:0909.2536 [hep-ph]} \BibitemShut
  {NoStop}%
\bibitem [{\citenamefont {Braun}(2012)}]{Braun:2011pp}%
  \BibitemOpen
  \bibfield  {author} {\bibinfo {author} {\bibfnamefont {J.}~\bibnamefont
  {Braun}},\ }\href {\doibase 10.1088/0954-3899/39/3/033001} {\bibfield
  {journal} {\bibinfo  {journal} {J.Phys.}\ }\textbf {\bibinfo {volume}
  {G39}},\ \bibinfo {pages} {033001} (\bibinfo {year} {2012})},\ \Eprint
  {http://arxiv.org/abs/1108.4449} {arXiv:1108.4449 [hep-ph]} \BibitemShut
  {NoStop}%
\bibitem [{\citenamefont {Roessner}\ \emph {et~al.}(2007)\citenamefont
  {Roessner}, \citenamefont {Ratti},\ and\ \citenamefont
  {Weise}}]{Roessner:2006xn}%
  \BibitemOpen
  \bibfield  {author} {\bibinfo {author} {\bibfnamefont {S.}~\bibnamefont
  {Roessner}}, \bibinfo {author} {\bibfnamefont {C.}~\bibnamefont {Ratti}}, \
  and\ \bibinfo {author} {\bibfnamefont {W.}~\bibnamefont {Weise}},\ }\href
  {\doibase 10.1103/PhysRevD.75.034007} {\bibfield  {journal} {\bibinfo
  {journal} {Phys. Rev.}\ }\textbf {\bibinfo {volume} {D75}},\ \bibinfo {pages}
  {034007} (\bibinfo {year} {2007})},\ \Eprint
  {http://arxiv.org/abs/hep-ph/0609281} {arXiv:hep-ph/0609281 [hep-ph]}
  \BibitemShut {NoStop}%
\bibitem [{\citenamefont {Schaefer}\ \emph {et~al.}(2007)\citenamefont
  {Schaefer}, \citenamefont {Pawlowski},\ and\ \citenamefont
  {Wambach}}]{Schaefer:2007pw}%
  \BibitemOpen
  \bibfield  {author} {\bibinfo {author} {\bibfnamefont {B.-J.}\ \bibnamefont
  {Schaefer}}, \bibinfo {author} {\bibfnamefont {J.~M.}\ \bibnamefont
  {Pawlowski}}, \ and\ \bibinfo {author} {\bibfnamefont {J.}~\bibnamefont
  {Wambach}},\ }\href {\doibase 10.1103/PhysRevD.76.074023} {\bibfield
  {journal} {\bibinfo  {journal} {Phys.Rev.}\ }\textbf {\bibinfo {volume}
  {D76}},\ \bibinfo {pages} {074023} (\bibinfo {year} {2007})},\ \Eprint
  {http://arxiv.org/abs/0704.3234} {arXiv:0704.3234 [hep-ph]} \BibitemShut
  {NoStop}%
\bibitem [{\citenamefont {Herbst}\ \emph {et~al.}(2011)\citenamefont {Herbst},
  \citenamefont {Pawlowski},\ and\ \citenamefont {Schaefer}}]{Herbst:2010rf}%
  \BibitemOpen
  \bibfield  {author} {\bibinfo {author} {\bibfnamefont {T.~K.}\ \bibnamefont
  {Herbst}}, \bibinfo {author} {\bibfnamefont {J.~M.}\ \bibnamefont
  {Pawlowski}}, \ and\ \bibinfo {author} {\bibfnamefont {B.-J.}\ \bibnamefont
  {Schaefer}},\ }\href {\doibase 10.1016/j.physletb.2010.12.003} {\bibfield
  {journal} {\bibinfo  {journal} {Phys.Lett.}\ }\textbf {\bibinfo {volume}
  {B696}},\ \bibinfo {pages} {58} (\bibinfo {year} {2011})},\ \Eprint
  {http://arxiv.org/abs/1008.0081} {arXiv:1008.0081 [hep-ph]} \BibitemShut
  {NoStop}%
\bibitem [{\citenamefont {Kobayashi}\ and\ \citenamefont
  {Maskawa}(1970)}]{Kobayashi:1970ji}%
  \BibitemOpen
  \bibfield  {author} {\bibinfo {author} {\bibfnamefont {M.}~\bibnamefont
  {Kobayashi}}\ and\ \bibinfo {author} {\bibfnamefont {T.}~\bibnamefont
  {Maskawa}},\ }\href {\doibase 10.1143/PTP.44.1422} {\bibfield  {journal}
  {\bibinfo  {journal} {Prog. Theor. Phys.}\ }\textbf {\bibinfo {volume}
  {44}},\ \bibinfo {pages} {1422} (\bibinfo {year} {1970})}\BibitemShut
  {NoStop}%
\bibitem [{\citenamefont {Skokov}\ \emph {et~al.}(2010)\citenamefont {Skokov},
  \citenamefont {Friman}, \citenamefont {Nakano}, \citenamefont {Redlich},\
  and\ \citenamefont {Schaefer}}]{Skokov:2010sf}%
  \BibitemOpen
  \bibfield  {author} {\bibinfo {author} {\bibfnamefont {V.}~\bibnamefont
  {Skokov}}, \bibinfo {author} {\bibfnamefont {B.}~\bibnamefont {Friman}},
  \bibinfo {author} {\bibfnamefont {E.}~\bibnamefont {Nakano}}, \bibinfo
  {author} {\bibfnamefont {K.}~\bibnamefont {Redlich}}, \ and\ \bibinfo
  {author} {\bibfnamefont {B.-J.}\ \bibnamefont {Schaefer}},\ }\href {\doibase
  10.1103/PhysRevD.82.034029} {\bibfield  {journal} {\bibinfo  {journal}
  {Phys.Rev.}\ }\textbf {\bibinfo {volume} {D82}},\ \bibinfo {pages} {034029}
  (\bibinfo {year} {2010})},\ \Eprint {http://arxiv.org/abs/1005.3166}
  {arXiv:1005.3166 [hep-ph]} \BibitemShut {NoStop}%
\bibitem [{\citenamefont {Haas}\ \emph {et~al.}(2013)\citenamefont {Haas},
  \citenamefont {Stiele}, \citenamefont {Braun}, \citenamefont {Pawlowski},\
  and\ \citenamefont {Schaffner-Bielich}}]{Haas:2013qwp}%
  \BibitemOpen
  \bibfield  {author} {\bibinfo {author} {\bibfnamefont {L.~M.}\ \bibnamefont
  {Haas}}, \bibinfo {author} {\bibfnamefont {R.}~\bibnamefont {Stiele}},
  \bibinfo {author} {\bibfnamefont {J.}~\bibnamefont {Braun}}, \bibinfo
  {author} {\bibfnamefont {J.~M.}\ \bibnamefont {Pawlowski}}, \ and\ \bibinfo
  {author} {\bibfnamefont {J.}~\bibnamefont {Schaffner-Bielich}},\ }\href
  {\doibase 10.1103/PhysRevD.87.076004} {\bibfield  {journal} {\bibinfo
  {journal} {Phys.Rev.}\ }\textbf {\bibinfo {volume} {D87}},\ \bibinfo {pages}
  {076004} (\bibinfo {year} {2013})},\ \Eprint {http://arxiv.org/abs/1302.1993}
  {arXiv:1302.1993 [hep-ph]} \BibitemShut {NoStop}%
\bibitem [{\citenamefont {Gies}\ and\ \citenamefont
  {Wetterich}(2002)}]{Gies:2001nw}%
  \BibitemOpen
  \bibfield  {author} {\bibinfo {author} {\bibfnamefont {H.}~\bibnamefont
  {Gies}}\ and\ \bibinfo {author} {\bibfnamefont {C.}~\bibnamefont
  {Wetterich}},\ }\href {\doibase 10.1103/PhysRevD.65.065001} {\bibfield
  {journal} {\bibinfo  {journal} {Phys.Rev.}\ }\textbf {\bibinfo {volume}
  {D65}},\ \bibinfo {pages} {065001} (\bibinfo {year} {2002})},\ \Eprint
  {http://arxiv.org/abs/hep-th/0107221} {arXiv:hep-th/0107221 [hep-th]}
  \BibitemShut {NoStop}%
\bibitem [{\citenamefont {Floerchinger}\ and\ \citenamefont
  {Wetterich}(2009)}]{Floerchinger:2009uf}%
  \BibitemOpen
  \bibfield  {author} {\bibinfo {author} {\bibfnamefont {S.}~\bibnamefont
  {Floerchinger}}\ and\ \bibinfo {author} {\bibfnamefont {C.}~\bibnamefont
  {Wetterich}},\ }\href {\doibase 10.1016/j.physletb.2009.09.014} {\bibfield
  {journal} {\bibinfo  {journal} {Phys.Lett.}\ }\textbf {\bibinfo {volume}
  {B680}},\ \bibinfo {pages} {371} (\bibinfo {year} {2009})},\ \Eprint
  {http://arxiv.org/abs/0905.0915} {arXiv:0905.0915 [hep-th]} \BibitemShut
  {NoStop}%
\bibitem [{\citenamefont {Braun}\ \emph {et~al.}(2014)\citenamefont {Braun},
  \citenamefont {Fister}, \citenamefont {Pawlowski},\ and\ \citenamefont
  {Rennecke}}]{Braun:2014ata}%
  \BibitemOpen
  \bibfield  {author} {\bibinfo {author} {\bibfnamefont {J.}~\bibnamefont
  {Braun}}, \bibinfo {author} {\bibfnamefont {L.}~\bibnamefont {Fister}},
  \bibinfo {author} {\bibfnamefont {J.~M.}\ \bibnamefont {Pawlowski}}, \ and\
  \bibinfo {author} {\bibfnamefont {F.}~\bibnamefont {Rennecke}},\ }\href@noop
  {} {\  (\bibinfo {year} {2014})},\ \Eprint {http://arxiv.org/abs/1412.1045}
  {arXiv:1412.1045 [hep-ph]} \BibitemShut {NoStop}%
\bibitem [{\citenamefont {Rennecke}(2015)}]{Rennecke:2015eba}%
  \BibitemOpen
  \bibfield  {author} {\bibinfo {author} {\bibfnamefont {F.}~\bibnamefont
  {Rennecke}},\ }\href {\doibase 10.1103/PhysRevD.92.076012} {\bibfield
  {journal} {\bibinfo  {journal} {Phys. Rev.}\ }\textbf {\bibinfo {volume}
  {D92}},\ \bibinfo {pages} {076012} (\bibinfo {year} {2015})},\ \Eprint
  {http://arxiv.org/abs/1504.03585} {arXiv:1504.03585 [hep-ph]} \BibitemShut
  {NoStop}%
\bibitem [{\citenamefont {Mueller}\ and\ \citenamefont
  {Pawlowski}(2015)}]{Mueller:2015fka}%
  \BibitemOpen
  \bibfield  {author} {\bibinfo {author} {\bibfnamefont {N.}~\bibnamefont
  {Mueller}}\ and\ \bibinfo {author} {\bibfnamefont {J.~M.}\ \bibnamefont
  {Pawlowski}},\ }\href {\doibase 10.1103/PhysRevD.91.116010} {\bibfield
  {journal} {\bibinfo  {journal} {Phys. Rev.}\ }\textbf {\bibinfo {volume}
  {D91}},\ \bibinfo {pages} {116010} (\bibinfo {year} {2015})},\ \Eprint
  {http://arxiv.org/abs/1502.08011} {arXiv:1502.08011 [hep-ph]} \BibitemShut
  {NoStop}%
\bibitem [{\citenamefont {Hashimoto}\ and\ \citenamefont
  {Izubuchi}(2008)}]{Hashimoto:2008xg}%
  \BibitemOpen
  \bibfield  {author} {\bibinfo {author} {\bibfnamefont {K.}~\bibnamefont
  {Hashimoto}}\ and\ \bibinfo {author} {\bibfnamefont {T.}~\bibnamefont
  {Izubuchi}},\ }\href {\doibase 10.1143/PTP.119.599} {\bibfield  {journal}
  {\bibinfo  {journal} {Prog. Theor. Phys.}\ }\textbf {\bibinfo {volume}
  {119}},\ \bibinfo {pages} {599} (\bibinfo {year} {2008})},\ \Eprint
  {http://arxiv.org/abs/0803.0186} {arXiv:0803.0186 [hep-lat]} \BibitemShut
  {NoStop}%
\bibitem [{\citenamefont {Pawlowski}\ and\ \citenamefont
  {Strodthoff}(2015)}]{Pawlowski:2015mia}%
  \BibitemOpen
  \bibfield  {author} {\bibinfo {author} {\bibfnamefont {J.~M.}\ \bibnamefont
  {Pawlowski}}\ and\ \bibinfo {author} {\bibfnamefont {N.}~\bibnamefont
  {Strodthoff}},\ }\href {\doibase 10.1103/PhysRevD.92.094009} {\bibfield
  {journal} {\bibinfo  {journal} {Phys. Rev.}\ }\textbf {\bibinfo {volume}
  {D92}},\ \bibinfo {pages} {094009} (\bibinfo {year} {2015})},\ \Eprint
  {http://arxiv.org/abs/1508.01160} {arXiv:1508.01160 [hep-ph]} \BibitemShut
  {NoStop}%
\end{thebibliography}

\onecolumngrid
\end{document}